\documentclass[fleqn,usenatbib]{mnras}

\usepackage{newtxtext,newtxmath}

\usepackage[T1]{fontenc}
\usepackage{ae,aecompl}
\usepackage{pdflscape}
\hypersetup{draft}

\usepackage{graphicx}	
\usepackage{amsmath}	
\usepackage{amssymb}	
\usepackage{gensymb}

\title[MUSE view of GRB 100316D]{The MUSE view of the host galaxy of GRB~100316D\thanks{Based on observations collected at ESO's Very Large Telescope under Prog-ID 092.D-0389(A) and 096.D-0786(A).}}

\author[L. Izzo et al.]{
L. Izzo,$^{1}$\thanks{E-mail: izzo@iaa.es}, C. C. Th\"one$^{1}$ , S. Schulze$^{2,3,4}$, A. Mehner$^{5}$, H. Flores$^{6}$, Z. Cano$^{1}$,
\newauthor\,  A. de Ugarte Postigo$^{1,7}$, D. A. Kann$^{1}$, R. Amor\'in$^{8,9}$, J. P. Anderson$^{5}$, F. E. Bauer$^{3,4,10}$,  
\newauthor\, K. Bensch$^{1}$, L. Christensen$^{7}$, S. Covino $^{11}$,  M. Della Valle$^{12}$,  J. P. U. Fynbo$^{7}$,
\newauthor\, P. Jakobsson$^{13}$, S. Klose$^{14}$, H. Kuncarayakti$^{15, 16}$, G. Leloudas$^{2}$,  B. Milvang-Jensen$^{7}$, 
\newauthor\, P. M\o{}ller$^{17}$, M. Puech$^{6}$, A. Rossi$^{18}$, R. S\'anchez-Ram\'irez$^{1}$, S. D. Vergani$^{6}$
\\
$^{1}$Instituto de Astrofisica de Andalucia (IAA-CSIC), Glorieta de la Astronomia s/n, E-18008 Granada, Spain \\
$^{2}$Department of Particle Physics and Astrophysics, Weizmann Institute of Science, Rehovot 7610001, Israel \\
$^{3}$Instituto de Astrof{\'{\i}}sica and Centro de Astroingenier{\'{\i}}a, Pontificia Universidad Cat{\'{o}}lica de Chile, Casilla 306, Santiago 22, Chile \\ 
$^{4}$Millennium Institute of Astrophysics (MAS), Nuncio Monse{\~{n}}or S{\'{o}}tero Sanz 100, Providencia, Santiago, Chile \\
$^{5}$European Southern Observatory, Alonso de C\'ordova 3107, Casilla 19, Santiago, Chile \\
$^{6}$GEPI , Observatoire de Paris, CNRS, University Paris Diderot, PSL Research University; 5 Place Jules Janssen, 92195 Meudon, France \\
$^{7}$Dark Cosmology center, Niels-Bohr-Institute, Univ. of Copenhagen, Juliane Maries Vej 30, 2100 Copenhagen, Denmark \\
$^{8}$Cavendish Laboratory, University of Cambridge, 19 JJ Thomson Avenue, Cambridge, CB3 0HE, UK \\
$^{9}$Kavli Institute for Cosmology, University of Cambridge, Madingley Road, Cambridge CB3 0HA, UK \\
$^{10}$Space Science Institute, 4750 Walnut Street, Suite 205, Boulder, Colorado 80301 \\
$^{11}$Osservatorio Astronomico di Brera, via Bianchi 46, 23807, Merate (LC), Italy \\
$^{12}$INAF, Osservatorio Astronomico di Capodimonte, salita Moiariello 16, I-80131, Napoli, Italy\\
$^{13}$ Center for Astrophysics and Cosmology, Science Institute, University of Iceland, Dunhagi 5, 107 Reykjavik, Iceland \\
$^{14}$Th\"uringer Landessternwarte Tautenburg, Sternwarte 5, 07778 Tautenburg, Germany \\
$^{15}$Finnish center for Astronomy with ESO (FINCA), University of Turku, V\"ais\"al\"antie 20, 21500 Piikki\"o, Finland\\
$^{16}$Tuorla Observatory, Department of Physics and Astronomy, University of Turku, V\"ais\"al\"antie 20, 21500 Piikki\"o, Finland\\
$^{17}$European Southern Observatory, Karl-Schwarzschildstrasse 2, 85748 Garching, Germany \\
$^{18}$INAF-IASF Bologna, Area della Ricerca CNR, via Gobetti 101, I-40129 Bologna, Italy \\
}

\date{Accepted XXX. Received YYY; in original form ZZZ}

\pubyear{2015}

\begin{document}
\label{firstpage}
\pagerange{\pageref{firstpage}--\pageref{lastpage}}
\maketitle

\begin{abstract}
The low distance, $z=0.0591$, of GRB 100316D and its association with SN 2010bh represent two important motivations for studying this host galaxy and the GRB's immediate environment with the Integral-Field Spectrographs like VLT/MUSE. Its large field-of-view allows us to create 2D maps of gas metallicity, ionization level, and the star-formation rate distribution maps, as well as to investigate the presence of possible host companions. The host is a late-type dwarf irregular galaxy with multiple star-forming regions and an extended central region with signatures of on-going shock interactions. The GRB site is characterized by the lowest metallicity, the highest star-formation rate and the youngest ($\sim$ 20-30 Myr) stellar population in the galaxy, which suggest a GRB progenitor stellar population with masses up to 20 -- 40 $M_{\odot}$.  We note that the GRB site has an offset of $\sim$660pc from the most luminous SF region in the host. The observed SF activity in this galaxy may have been triggered by a relatively recent gravitational encounter between the host and a small undetected ($L_{H\alpha} \leq 10^{36}$ erg/s) companion.
\end{abstract}




\section{Introduction}

Long gamma ray bursts (GRBs) are among the most luminous explosions in the Universe \citep[see][for a review on their phenomenology]{Gehrels2009}, originating in catastrophic events such as the collapse of very massive stars \citep[the \textit{collapsar} scenario,][]{Woosley1993,MacFadyenWoosley}. Their association with broad-lined Type Ic supernovae (SNe) \citep{Galama1998,Hjorth2003,Pian2006} and occurrence in the brightest star-forming regions of their host galaxies \citep{Bloom2002,Fruchter2006,Kelly2008,Blanchard2016,Lyman2017} support this model. 

The discovery of the broad-lined type Ic supernova 1998bw in the error box of GRB 980425 gave a flying head start to the hypothesis that long GRBs are connected to the collapse of very massive stars \citep{Paczynski1998,Galama1998}. Since then nearly 50 GRB-SNe have been  detected \citep{Cano2017}, but only a dozen of these have a spectroscopic confirmation of the SN signatures \citep{Modjaz2016}. While the properties of most SNe associated with GRBs are very similar to those of SN 1998bw \citep[for a review see ][]{Cano2017}, the properties of the GRB itself show a rich diversity, such as burst durations between 2 and $>10\,000$ s and an isotropic gamma-ray luminosity between $10^{46}$ and $> 10^{52}$ erg/s. To explain this diversity, \cite{Bromberg2011} suggested that GRBs with an isotropic gamma-ray luminosity of $L_{\rm iso}<10^{48.5}~{\rm erg\,s}^{-1}$ have a weak jet that is either chocked or barely manages to penetrate the stellar envelope \citep[see also][]{Bromberg2011b}. Their $\gamma$-ray emission is powered by high-energy emission from the shock break-out of the progenitor star. In contrast, high-luminosity GRBs with $L_{\rm iso}>10^{49.5}~{\rm erg\,s}^{-1}$ are powered by collimated ultra-relativistic jets that successfully penetrate the stellar envelope. These distinct interpretations and the different rates for low- and high-L GRBs \citep{Pian2006, Chapman2007, Guetta2007, Liang2007, Virgili2009, Wanderman2010} suggested that the two sub-classes are connected to distinct populations of massive stars.

There are few remarkable outliers from these two sub-classes: \citet{Thoene2011} detected an exceptionally faint GRB-SN accompanying the peculiar GRB 101225A with a peak $V$-band luminosity 2.5 mag fainter than SN 1998bw \citep[but see also][]{Levan2014}. On the other extreme, \citet{Greiner2015} reported an exceptionally bright GRB-SN accompanying GRB 111209A \citep[see also][]{Kann2016} that is similar to hydrogen-poor super-luminous supernovae \citep[SLSNe, for a review see][]{GalYam2012}. Furthermore, no SN was found for three out of 24 cases of nearby long GRBs ($z<0.5$), to limits several magnitudes deeper than any GRB-SN \citep{Fynbo2006, GalYam2006, DellaValle2006, Kann2011, Michalowski2016, Tanga2017}. \citet{Fynbo2006} suggested that the kinetic energy of the SN ejecta could have been too low to avoid fall-back on the forming black hole. Another possibility is that the collapse only produced small amounts of nickel and hence a SN too faint to be detected \citep{Heger2003, Fryer2006, Tominaga2007}.

This diversity could imply the existence of multiple progenitor channels for long GRBs. Despite the success of spectroscopic and photometric studies on GRB host galaxies \citep{LeFloch2003,Christensen2004,Savaglio2009,Rossi2012,Hjorth2012,Perley2013,Kruhler2015,Schulze2015,Vergani2017}, the lack of spatial information only provides spatially averaged host properties. It is not clear how and if these integrated measurements are representative for the conditions at the GRB explosion sites and how star-formation varies across the host galaxies. One way to investigate this potential diversity is by zooming in on the explosion sites to examine their properties. Up to now, integral field unit (IFU) spectroscopy was only performed for the host galaxies of GRBs 980425 \citep{Christensen2008,Kruhler2017} and 060505 \citep{Thoene2014}. These works have indeed shown that some physical properties vary between individual \ion{H}{II} regions. For this reason it is crucial to investigate the immediate GRB environment at high spatial resolution to obtain accurate information on the physical properties of the region where the GRB progenitor spent its entire evolution \citep{Modjaz2011} and comparing these properties to other star-forming regions in the host galaxy. For this reason we carried out a systematic survey of all extended GRB host galaxies at $z<0.3$ using the new wide-field medium-resolution Multi-Unit Spectroscopic Explorer (MUSE; \citealt{Bacon2010}) at the ESO Very Large Telescope (VLT). The scope of this survey is to examine \textit{i}) the conditions of the GRB/SN explosion sites to pin down the progenitor population(s), \textit{ii}) how star-formation varies across their host galaxies, and \textit{iii}) how their galaxy environment affects the star-formation process.

In this work we present a study with MUSE of the nearby ($z = 0.059$) host galaxy of GRB~100316D/SN~2010bh, a sub-energetic burst ($E_{\textrm{iso}} \geq 5.9 \times 10^{49}$ erg), associated with the SN 2010bh \citep{Olivares2012, Bufano2012} and characterized by the presence of a possible thermal component in its X-ray emission \citep{Starling2011}. The host galaxy of GRB~100316D is a blue galaxy with an extended (diameter $\sim$ 12$''$) and disturbed morphology. Imaging obtained with the \textit{Hubble Space Telescope (HST)} shows the presence of a large number of giant \ion{H}{II} regions, with the GRB located close to the brightest of these knots \citep{Starling2011}. We complement the analysis with an additional dataset obtained with the IFU FLAMES/GIRAFFE at the VLT, whose field of view covers a limited region of the host containing the GRB site, but its higher spectral resolution provides more detailed information on the velocity distribution in the galaxy. 

We discuss the IFU observations in Section 2. In Section 3 we present the procedure used for the analysis of the data cubes, the identification of bright \ion{H}{II} regions and the results obtained from their analysis. We estimate different physical properties such as the extinction, the metallicity, the ionization and the SFR for all these regions and the GRB site, identified by using the \textit{HST} astrometrically-corrected position \citep{Starling2011}. In Section 4 we discuss the importance of spatially-resolved observations of GRB host galaxies comparing the results obtained from single galaxy regions with an integrated spectrum of the entire host galaxy, while in Section 5 we present the results of the search for possible host galaxy companions. In Section 6 we present a kinematic analysis of the entire host galaxy, and in the last section we discuss our results. We adopt a standard $\Lambda$CDM cosmological model with $H_0$ = 70 km/s/Mpc, $\Omega_m$ = 0.3 and $\Omega_{\Lambda}$ = 0.7. We also adopt as the solar metallicity the value of 12 + log(\ion{O}/\ion{H}) = 8.69 \citep{Asplund2009}.

\begin{figure*}
	\includegraphics[width=8.8cm]{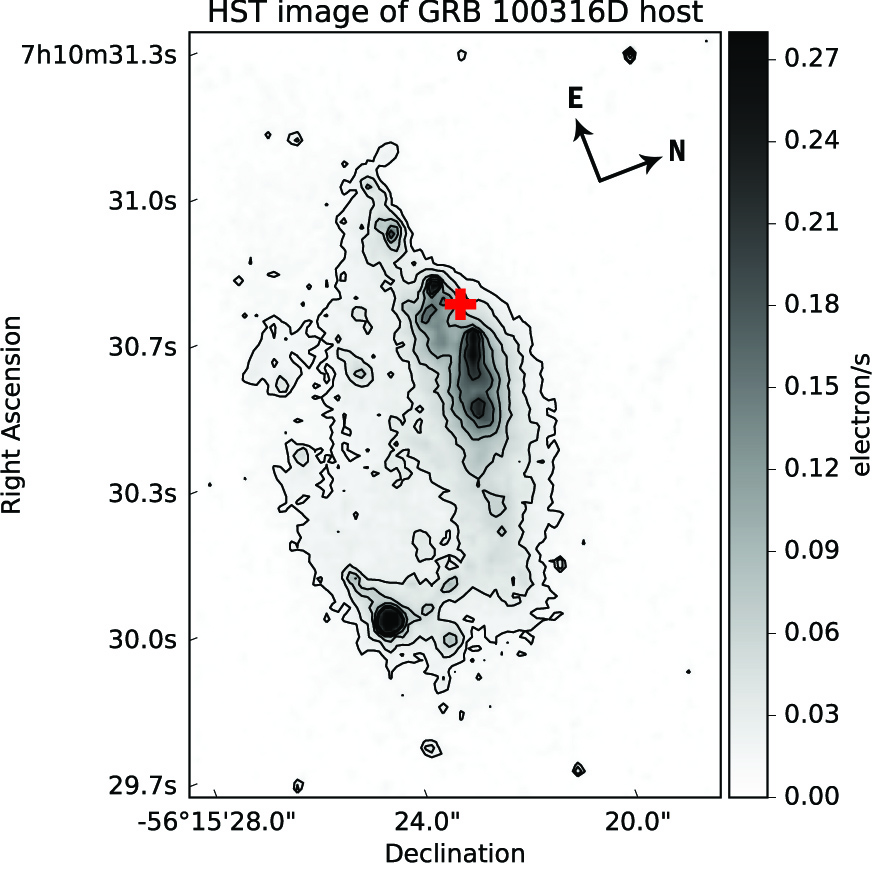}
    \includegraphics[width=8.8cm]{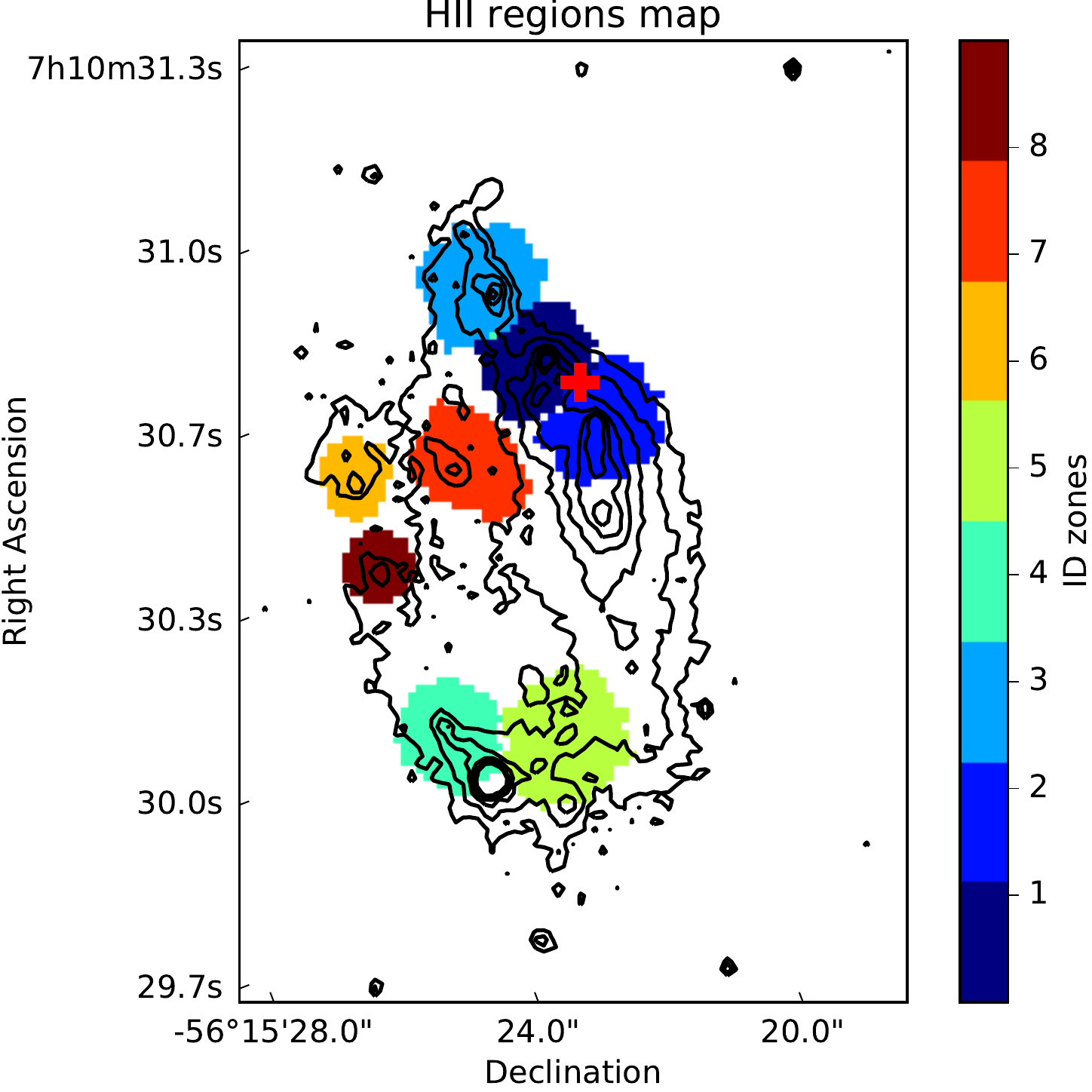}
    \caption{\textit{(Left panel)} \textit{HST}/WFC3 image ($9'' \times 13''$) of the host galaxy of GRB~100316D obtained in the filter F625W, which also covers the H$\alpha$ emission line. The contours from this image are used as a reference for all the maps presented in this work. The red cross corresponds to the \textit{HST} astrometric location of the GRB while black contours correspond to galaxy isophotes. The North-East orientation is the same for all the maps shown in this paper, with the exception of Fig. \ref{fig:no8b}. (\textit{Right panel}) The location and extension of the eight \ion{H}{II} regions found by an automatic \ion{H}{II} region search algorithm (see description in the text). The bright point-like source located in between the \ion{H}{II}-4 and \ion{H}{II}-5 regions is a foreground star.}
    \label{fig:no1}
\end{figure*}

\section{Observations}

GRB~100316D was observed at the Very Large Telescope with MUSE \citep{Bacon2010} on December 9, 2015 (Program ID 096.D-0786, PI S. Schulze). MUSE is an integral field spectrograph, composed of 24 integral field units that sample a contiguous 1\arcmin$\times$1\arcmin\ field of view in the wide-field mode WFM-NOAO-N used in this paper. The instrument covers the wavelength  region from 4\,800--9\,300~\AA\ with a spectral resolving power of $R\sim1\,600$ (blue) to $R\sim3\,600$ (red) and a spatial sampling of 0\farcs2. Three exposures of 1200~s each were obtained with GRB~100316D located in the center of the field of view. The instrument was rotated by 90\degree\ between exposures to average out the pattern of the slicers and channels in the reduction process. The data were reduced using version 1.2.1 of the MUSE ESO standard pipeline \citep{2012SPIE.8451E..0BW}. Bias, arc, and flat-field master calibration files were created using a default set of calibration exposures. These include an illumination-correction flat field taken within one hour of the science observations to correct for temperature variation in the illumination pattern of the slices. All data cubes were sampled to $0\farcs2\times0\farcs2\times1.25$ \AA\, and the three exposures were combined. After reducing the data, we removed sky lines with the Zurich Atmosphere Purge (ZAP) software package \citep{Soto2016a} on the combined image for each spectral pixel (spaxel). The final MUSE data cube has a nominal resolution of 1\farcs1 (due to the seeing conditions) which corresponds to a distance of 1.3 kpc at the redshift of the host galaxy. 

For the kinematic analysis of the host we furthermore used data from FLAMES/GIRAFFE at the VLT in ARGUS mode (Program ID 092.D-0389, PI C. C. Th\"one). Observations were made in the LR6 set-up (wavelength range 6438 -- 7184 \AA\,, resolving power R = 13500), with a total integration time of $9\times15$ min. Data reduction was done using the standard ESO pipeline (version 2.11) without the sky subtraction options. In order to verify the fiber-to-fiber wavelength calibration, we checked the wavelength of two skylines in the ARGUS data cube.  Data cubes for each single exposure were constructed using IRAF and IDL dedicated software \citep{Flores2006,Yang2008} and then combined to a final datacube.

\begin{figure}
	\includegraphics[width=9.2cm,height=6.3cm]{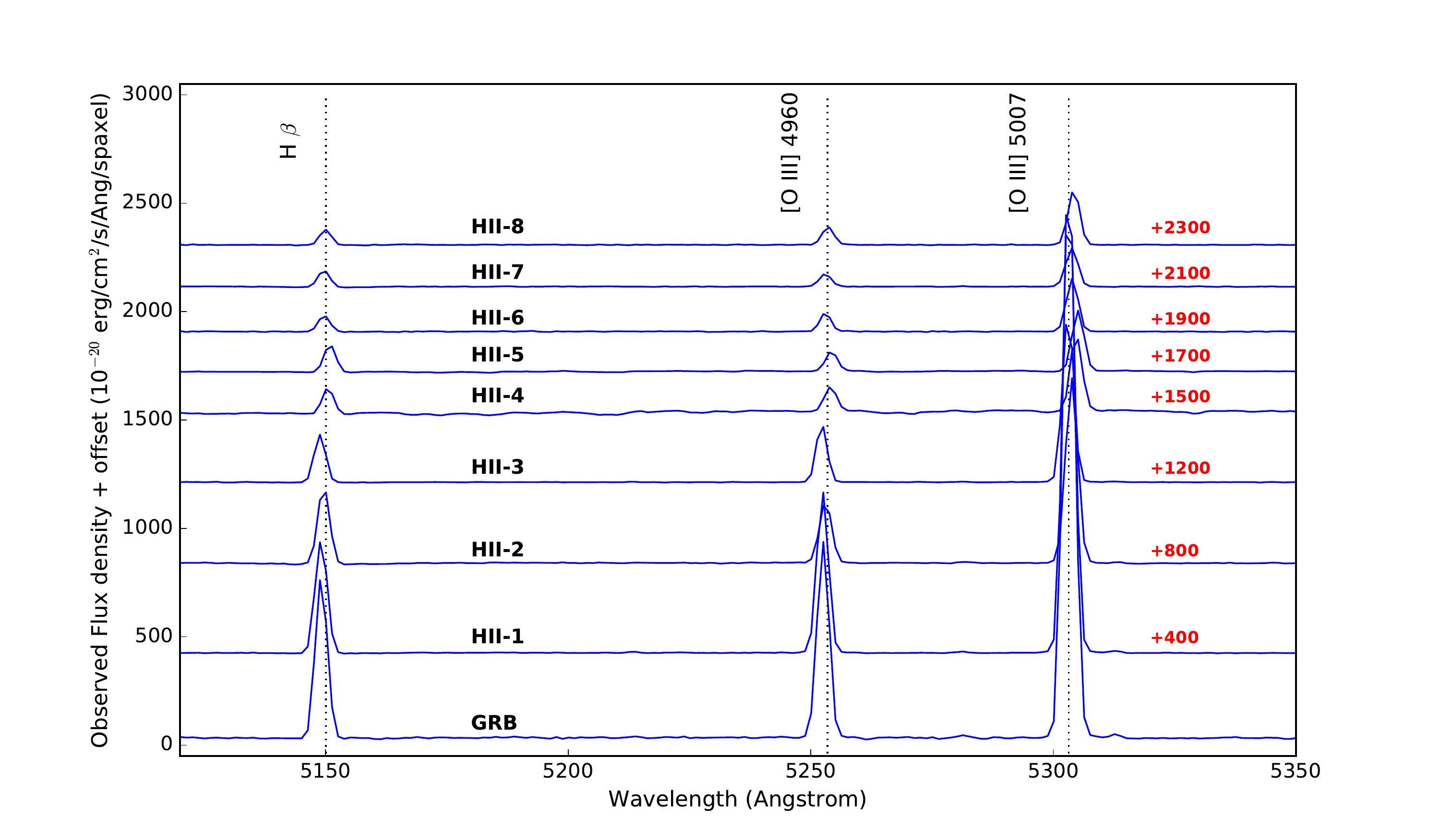}
    \caption{Spaxel-averaged spectra in the wavelength interval, (5110-5350 \AA\,), that cover the range of the H$\beta$ and [\ion{O}{III}] $\lambda\lambda$4959,5007 emission lines identified, for the different \ion{H}{II} regions and GRB site. The spectra are offset by an arbitrary value for the sake of clarity.}
    \label{fig:no2}
\end{figure}

\section{Resolved Properties}

The host of GRB~100316D is a low-mass ($\textrm{log}\, M_*/M_{\odot}$ = 8.93) blue, irregular galaxy \citep{Michalowski2015}, characterized by the presence of several \ion{H}{II} regions (see Fig. \ref{fig:no1}). Their distribution suggests an on-going and very active star-bursting phase. From the MUSE data we created maps of emission lines by summing the flux of 15 pixels in wavelength around the center of the corresponding red-shifted emission line in each spaxel, which take into account the velocity distribution of the line in the host given the spectral sampling of MUSE (1.25 \AA). The continuum was subtracted taking the average value of 10 pixels red- and bluewards with respect to the emission line center and selected to be free of other features, such as emission or telluric absorption lines. 

We identified individual \ion{H}{II} regions in the galaxy using a modified \textsc{python} version of the \ion{H}{II} explorer algorithm \citep{Sanchez2012}. This code automatically identifies regions in the H$\alpha$ map with a flux larger than an assigned value, in our case $F_{pk} = 10^{-17}$ erg/cm$^2$/s corresponding to an H$\alpha$ luminosity equal to or larger than 10$^{38}$ erg/s for the GRB host, a value that divides large giant \ion{H}{II} regions (like 30 Dor, with an H$\alpha$ luminosity of 5 $\times$ 10$^{39}$ erg/s) from normal ones \citep{Kennicutt1989b}. The algorithm includes for each region all spaxels with an H$\alpha$ flux larger than a fixed threshold value $F_{th}$ = 10$^{-18}$ erg/cm$^2$/s, and finally assumes a maximum size of $\sim$ 1 kpc for an \ion{H}{II} region. This means that only spaxels whose distance from the corresponding region peak was not more than 5 spaxels are included. The choice of 1 kpc for the maximum size of a \ion{H}{II} region is based on 1) a seeing of (1$''$) at the epoch of observations, which for a redshift of $z=0.0591$ corresponds to a scale of 1.1$''$/kpc; 2) the fact that this is similar to the size of large \ion{H}{II} regions in external galaxies, as for example NGC5641 and NGC5471 in the M101 galaxy \citep{Kennicutt1984}.  The algorithm then identifies \ion{H}{II} regions following the above constraints and then extracts a corresponding integrated spectrum. We also extracted a spectrum of the astrometrically determined GRB position and the eight immediately adjacent spaxels. We note that the region corresponding to the GRB spectrum falls in between two distinct \ion{H}{II} regions that are also the brightest ones in the galaxy: the apparent distance between the GRB location and the peak of the \ion{H}{II}-1 region is 0.6$''$, which at the GRB redshift corresponds to a projected distance of 660 pc. Fig. \ref{fig:no1} shows the \textit{HST}/WFC3 (PI D. Bersier) F625W-filter image and the eight \ion{H}{II} regions found with the algorithm described above, together with the exact location of the GRB. The coordinates of the emission peak for the \ion{H}{II} regions and of the GRB location are reported in Table \ref{tab:no0}.
    
\begin{table}
	\centering
	\caption{Coordinates of the emission peaks for the \ion{H}{II} regions and the GRB location (see Fig.\ref{fig:no1}.}
	\label{tab:no0}
	\begin{tabular}{lcc}
\hline
 Region        & RA      & Dec.                       \\
  & (J2000) & (J2000) \\
\hline
 \ion{H}{II}-1 & 07h 10m 30.01s & -56d 15m 23.3s \\
 \ion{H}{II}-2 & 07h 10m 30.02s & -56d 15m 22.5s  \\
 \ion{H}{II}-3 & 07h 10m 30.02s & -56d 15m 24.2s \\
 \ion{H}{II}-4 & 07h 10m 29.64s & -56d 15m 21.3s \\
 \ion{H}{II}-5 & 07h 10m 29.74s & -56d 15m 20.5s \\
 \ion{H}{II}-6 & 07h 10m 29.76s & -56d 15m 23.6s \\
 \ion{H}{II}-7 & 07h 10m 29.86s & -56d 15m 23.1s  \\
 \ion{H}{II}-8 & 07h 10m 29.71s & -56d 15m 22.9s \\
 GRB & 07h 10m 30.01s & -56d 15m 22.8s \\
\hline
\end{tabular}
\end{table}

\begin{figure*}
	\includegraphics[width=8.8cm]{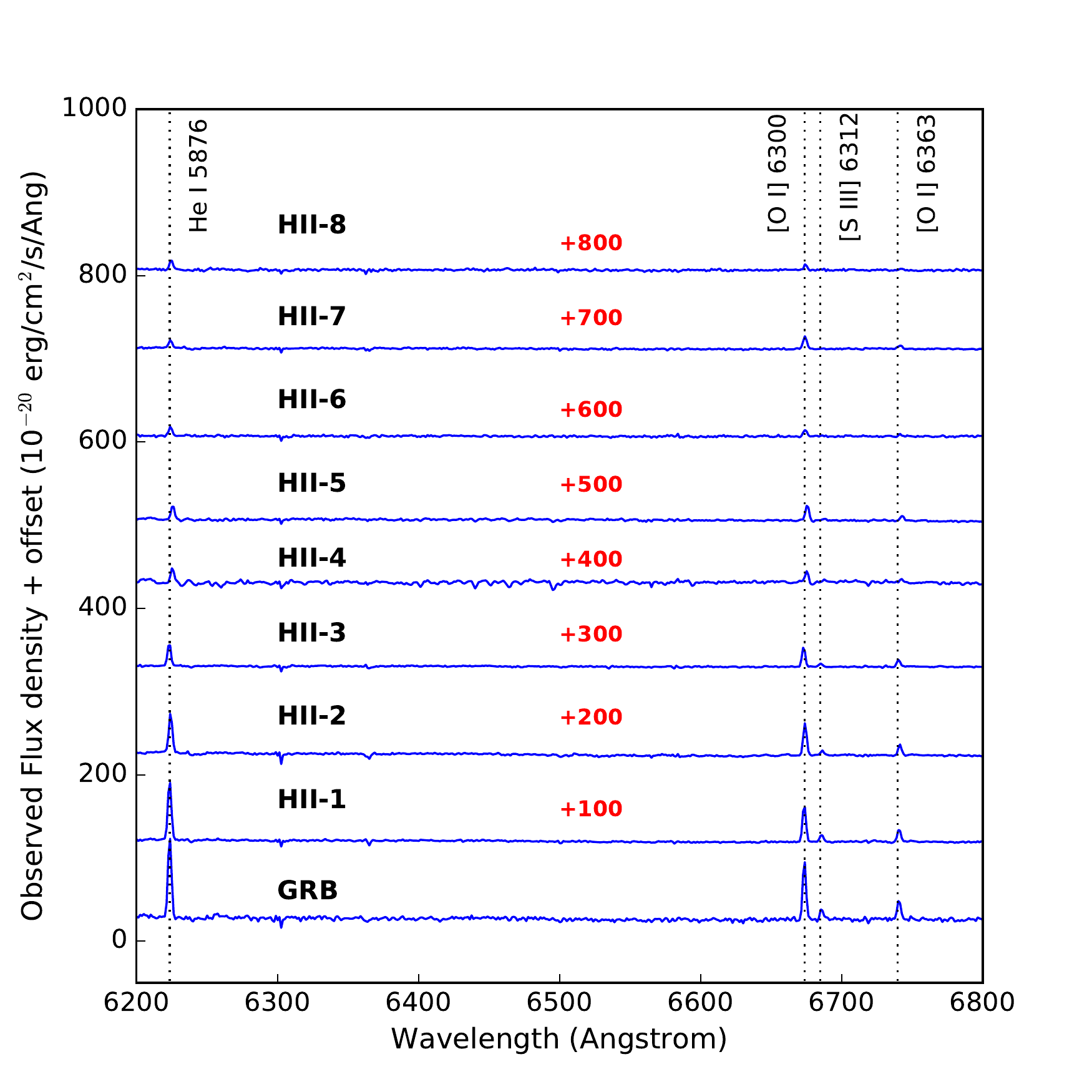}
    \includegraphics[width=8.8cm]{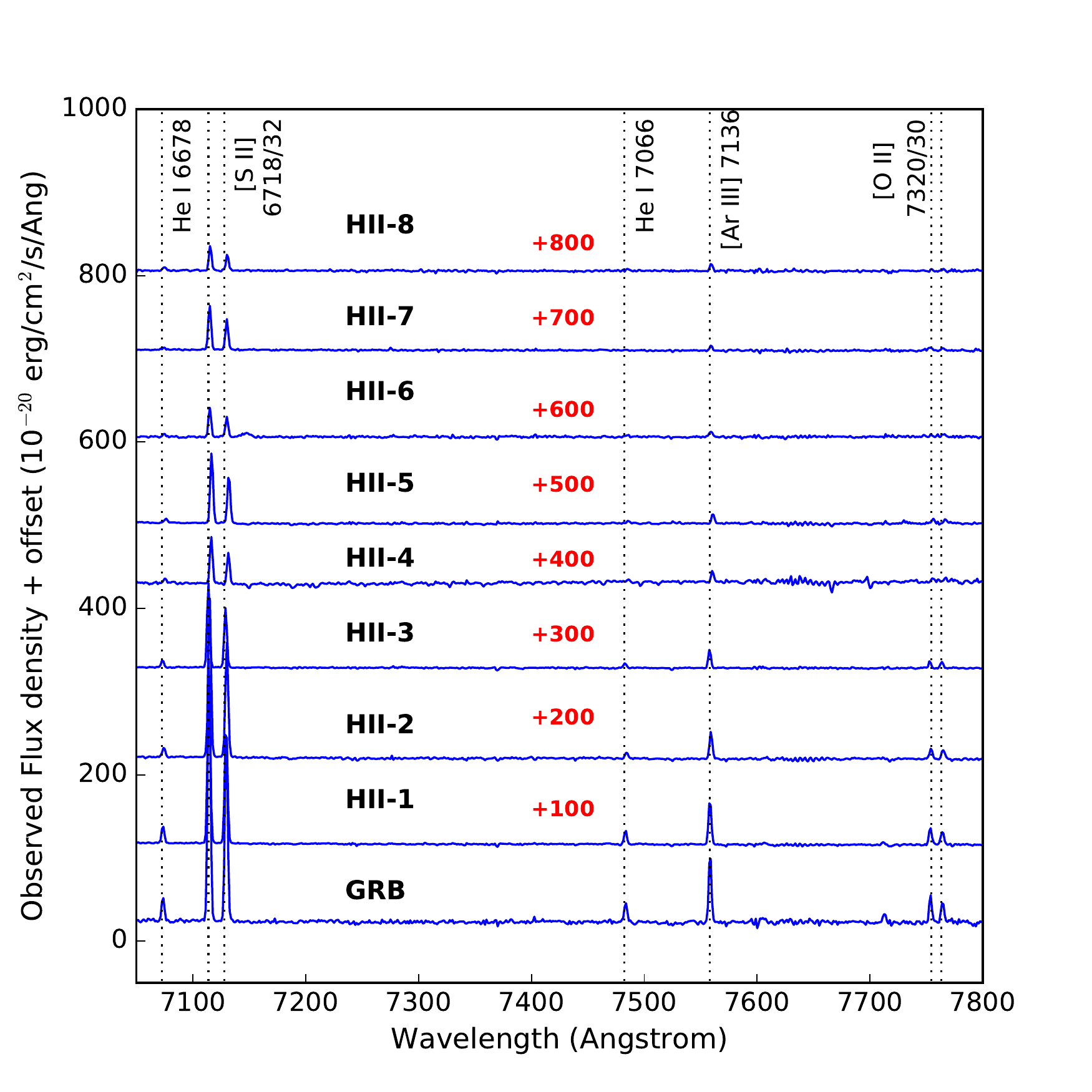}
    \caption{Selected regions of the spaxel-averaged spectra of the \ion{H}{II} regions and GRB site. The left panel shows the range 6200--6800 \AA\, characterized by the presence of the \ion{He}{I} $\lambda$~5876 and [\ion{O}{I}] $\lambda\lambda$~6300,6363 lines, while the right panel shows the range 7050-7800 \AA\, covering the [\ion{S}{II}] and the [\ion{O}{II}] doublets, and other lines such as \ion{He}{I} $\lambda\lambda$~6678,7066 and [\ion{Ar}{III}] $\lambda$~7136. The spectra of different regions are offset using arbitrary constants.}
    \label{fig:no2b}
\end{figure*}

In all \ion{H}{II} region spectra we detect the strong emission lines of H$\alpha$, H$\beta$, the doublets of [\ion{O}{III}] $\lambda\lambda$~4959,5007, [\ion{N}{II}] $\lambda\lambda$~6550/86, [\ion{S}{II}] $\lambda\lambda$~6718/32 , and [\ion{O}{II}] $\lambda\lambda$~7320/30. Given the wavelength range of MUSE and redshift of the host galaxy of $z = 0.0591$ we can not observe the doublet [\ion{O}{II}] $\lambda\lambda$~3727/29. Line fluxes and properties of the integrated spectra for the GRB and for each \ion{H}{II} region are reported in Tables \ref{tab:no1}, \ref{tab:no2}, and \ref{tab:no2b}, see also Fig. \ref{fig:no2}. We note that source A reported in \citet{Starling2011} corresponds to the region \ion{H}{II}-3  described in this work: the same lines are identified, with the exception of the [\ion{O}{II}] $\lambda\lambda$~3727/29 doublet and the [\ion{O}{III}] $\lambda$~4363 due to the lower wavelength limit of the MUSE data ($\lambda_{\textrm{min}}$ = 4750 \AA). Given the high sensitivity of MUSE, we also identify several transitions of \ion{He}{I} $\lambda$~4922, $\lambda$~5016, $\lambda$~6678, $\lambda$ 7066, as well as [\ion{N}{I}] $\lambda$~5199, [\ion{Ar}{III}] $\lambda$~7136, $\lambda$~7751, [\ion{S}{III}] $\lambda$~6312 and [\ion{Fe}{III}] $\lambda$~4659, $\lambda$~4987, see also Fig. \ref{fig:no2b}. Fluxes for these lines are reported in Table \ref{tab:app1}.

All the maps presented in this paper have been reprojected onto the World Coordinate system (WCS) of the \textit{HST}/WFC3 image, after an appropriate astrometrical correction of the MUSE and FLAMES cubes using the \texttt{astropy} python package \citep{astropy}, whose set of libraries has been extensively used in this work.

\begin{table*}
	\centering
	\caption{Extinction-corrected fluxes measured through Gaussian fits of the corresponding emission lines in units of 10$^{-18}$ erg/cm$^2$/s, see also Figs. \ref{fig:no1},\ref{fig:no2}. Balmer lines are corrected for the stellar absorption in the continuum. A table with the fluxes of weaker lines can be found in the appendix.}
	\label{tab:no1}
	\begin{tabular}{lcccccccc}
\hline
 Region   & H$\beta$            & [\ion{O}{III}]    & [\ion{O}{III}]   & H$\alpha$              & [\ion{N}{II}]   & [\ion{S}{II}]    & [\ion{S}{II}]    & [\ion{O}{II}]   \\
     &            & $\lambda$~4959   & $\lambda$~5007   &              &  $\lambda$~6584   &  $\lambda$~6718   &  $\lambda$~6732   &$\lambda\lambda$~7320/30   \\
\hline
 HII-1    & 1859.31$\pm$6.76 & 2380.01$\pm$7.52      & 6921.98$\pm$11.04     & 6562.62$\pm$10.26 & 384.90$\pm$4.49       & 683.52$\pm$5.00       & 491.43$\pm$4.65      & 152.79$\pm$3.54         \\
 HII-2    & 1334.37$\pm$6.74 & 955.17$\pm$6.82       & 4928.77$\pm$8.80       & 4349.54$\pm$9.59 & 459.45$\pm$5.57      & 666.83$\pm$5.87      & 480.25$\pm$5.55      & 79.97$\pm$4.43          \\
 HII-3    & 725.85$\pm$4.37  & 773.81$\pm$4.67       & 2669.88$\pm$6.59      & 2473.45$\pm$6.362  & 158.05$\pm$3.03      & 310.36$\pm$3.41      & 222.43$\pm$3.18      & 56.65$\pm$2.44          \\
 HII-4    & 196.12$\pm$4.79  & 322.07$\pm$5.34       & 777.07$\pm$6.42       & 1194.12$\pm$7.14 & 76.53$\pm$5.34       & 162.67$\pm$5.52      & 89.71$\pm$5.36       & 27.28$\pm$5.26          \\
 HII-5    & 519.21$\pm$5.08  & 340.84$\pm$5.15       & 2015.25$\pm$6.27      & 1861.21$\pm$7.16 & 195.30$\pm$4.73       & 336.98$\pm$4.94      & 233.76$\pm$4.72      & 51.85$\pm$4.10           \\
 HII-6    & 92.97$\pm$1.78   & 95.67$\pm$1.93        & 354.96$\pm$2.51       & 310.53$\pm$2.47  & 22.73$\pm$1.39       & 39.02$\pm$1.48       & 28.10$\pm$1.43        & 11.50$\pm$1.24           \\
 HII-7    & 285.87$\pm$2.91  & 122.05$\pm$3.00        & 927.99$\pm$3.65       & 570.49$\pm$3.81  & 55.29$\pm$2.41       & 118.04$\pm$2.66      & 81.60$\pm$2.52        & 29.83$\pm$2.08          \\
 HII-8    & 78.15$\pm$1.86   & 108.57$\pm$2.06       & 352.66$\pm$2.69       & 352.76$\pm$2.62  & 22.41$\pm$1.42       & 36.25$\pm$1.53       & 25.20$\pm$1.46        & 3.9$\pm$1.20             \\
 GRB      & 1716.70$\pm$6.47 & 1982.19$\pm$6.95      & 6552.81$\pm$10.20      & 6307.71$\pm$9.99 & 433.12$\pm$4.47      & 735.16$\pm$4.95      & 534.05$\pm$4.60       & 137.14$\pm$3.42         \\
\hline
\end{tabular}
\end{table*}

\subsection{Extinction}

\begin{figure}
	\includegraphics[width=\columnwidth]{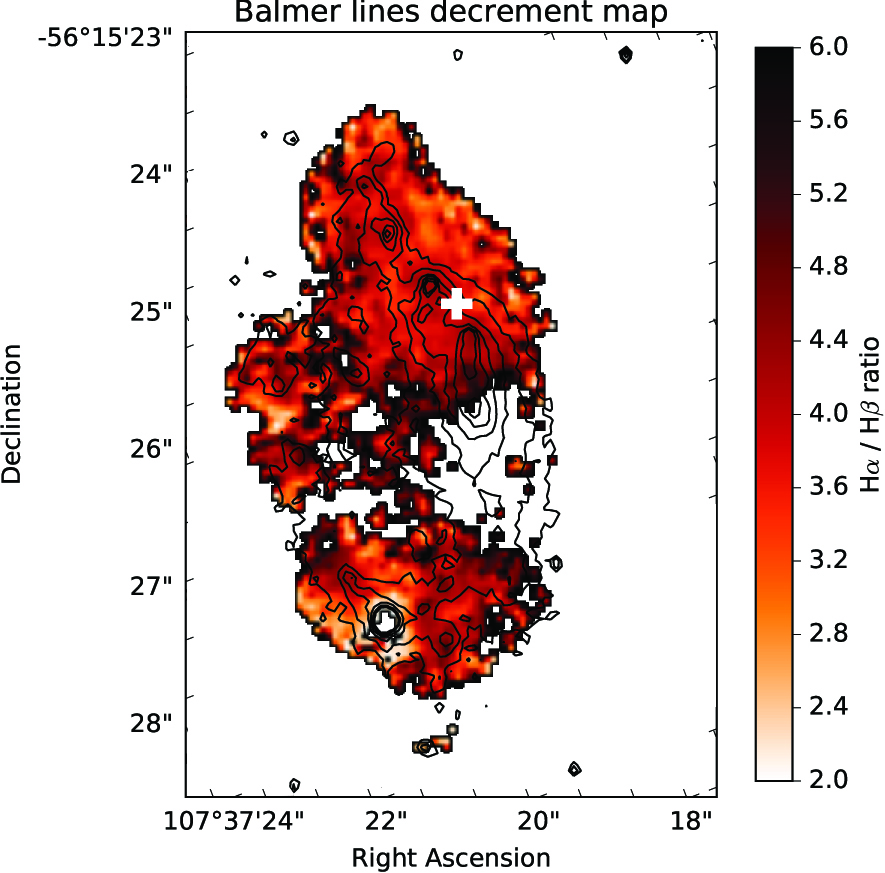}
    \caption{Balmer decrement obtained using the line fluxes ratio of the Balmer H$\alpha$ and H$\beta$ lines for case B recombination. The GRB location is shown as a white cross.}
    \label{fig:no2a}
\end{figure}

We computed the host galaxy internal extinction by first determining the effective absorption curve $\tau(\lambda)$ for each region (see also Fig. \ref{fig:no2a}) using the Balmer-line decrement (H$\alpha$/H$\beta$) for case B recombination \citep{Osterbrock} and the intrinsic color excess $E(B-V)$ by assuming $R_V$ = 4.5 for the GRB host galaxy, as found by \citet{Wiersema2011}. We also correct the Balmer emission lines in all \ion{H}{II} regions, as well as the GRB one, for the observed underlying stellar absorption component (see additional details in Section 3.5) to the corresponding spectral regions and correcting the observed flux for this stellar absorption, before measuring the flux and the equivalent width (EW) of the emission lines. The extinction for the region corresponding to the GRB site has a value of $E(B-V) = 0.42 \pm 0.01$ mag. For completeness we also compute the extinction using a Milky Way (MW) ($R_V$ = 3.1) as well as a Small Magellanic Cloud (SMC)  extinction law ($R_V$= 2.76), which is usually used for faint dwarf galaxies and obtain values of $E(B-V)_{GRB} = 0.61 \pm 0.01$ for the MW case and $E(B-V)_{GRB} = 0.69 \pm 0.01$ for the SMC case.

\subsection{Metallicity}

\begin{table*}
	\centering
	\caption{Physical properties derived from the integrated spectra of the \ion{H}{II} regions. U refers to the ionization parameter as defined in \citet{Diaz2000}. EWs of emission lines are reported in the observer frame.}
	\label{tab:no2}
	\begin{tabular}{lcccccccc}
\hline
 Region   & log U     & log U$_{\textrm{PM}}$  & SFR             & sSFR/(L/L$^*$)          & $E(B-V)$      & EW(H$\alpha$)        & EW(H$\beta$)      & EW([\ion{O}{III}])   \\
     &     &    & (M$_{\odot}$ yr$^{-1}$)             & (M$_{\odot}$ yr$^{-1}$)          & (mag)      & \AA\,       & \AA\,      & \AA\,   \\
\hline
 HII-1    & -2.66$\pm$0.01 & -2.81$\pm$0.01 & 0.463$\pm$0.001 & 0.36$\pm$0.03 & 0.39$\pm$0.01  & -324.41$\pm$22.51 & -70.34$\pm$10.14 & -248.26$\pm$19.70    \\
 HII-2    & -2.76$\pm$0.02 & -2.95$\pm$0.03 & 0.33$\pm$0.001  & 0.18$\pm$0.03 & 0.38$\pm$0.01  & -132.59$\pm$14.39 & -32.62$\pm$6.78  & -65.44$\pm$10.11    \\
 HII-3    & -2.74$\pm$0.02 & -2.86$\pm$0.02 & 0.179$\pm$0.001   & 0.29$\pm$0.02 & 0.38$\pm$0.01  & -257.04$\pm$20.04  & -55.44$\pm$9.17  & -152.14$\pm$15.42   \\
 HII-4    & -3.03$\pm$0.07 & -2.91$\pm$0.05 & 0.052$\pm$0.001   & 0.04$\pm$0.02 & 0.47$\pm$0.01 & -25.91$\pm$6.36   & -21.21$\pm$3.79  & -22.45$\pm$5.92     \\
 HII-5    & -2.91$\pm$0.03 & -2.99$\pm$0.01 & 0.135$\pm$0.001   & 0.11$\pm$0.02 & 0.44$\pm$0.01  & -62.41$\pm$9.88   & -19.03$\pm$5.04  & -28.26$\pm$6.65     \\
 HII-6    & -2.72$\pm$0.07 & -2.98$\pm$0.01 & 0.023$\pm$0.001   & 0.17$\pm$0.02 & 0.39$\pm$0.01 & -146.16$\pm$15.11 & -32.60$\pm$6.45   & -77.92$\pm$11.03    \\
 HII-7    & -2.67$\pm$0.04 & -2.91$\pm$0.05 & 0.062$\pm$0.001   & 0.14$\pm$0.02 & 0.22$\pm$0.01  & -79.86$\pm$11.17  & -29.44$\pm$4.81  & -33.48$\pm$7.23     \\
 HII-8    & -2.78$\pm$0.08 & -2.82$\pm$0.01 & 0.024$\pm$0.001   & 0.16$\pm$0.02 & 0.63$\pm$0.02 & -150.02$\pm$15.31  & -26.25$\pm$6.40   & -87.60$\pm$11.70      \\
 GRB      & -2.74$\pm$0.01 & -2.82$\pm$0.01 & 0.439$\pm$0.001 & 0.38$\pm$0.03 & 0.42$\pm$0.01  & -330.86$\pm$22.74 & -75.35$\pm$10.44 & -229.17$\pm$18.92   \\
\hline
\end{tabular}
\end{table*}

\begin{figure*}
	\includegraphics[width=0.65\columnwidth]{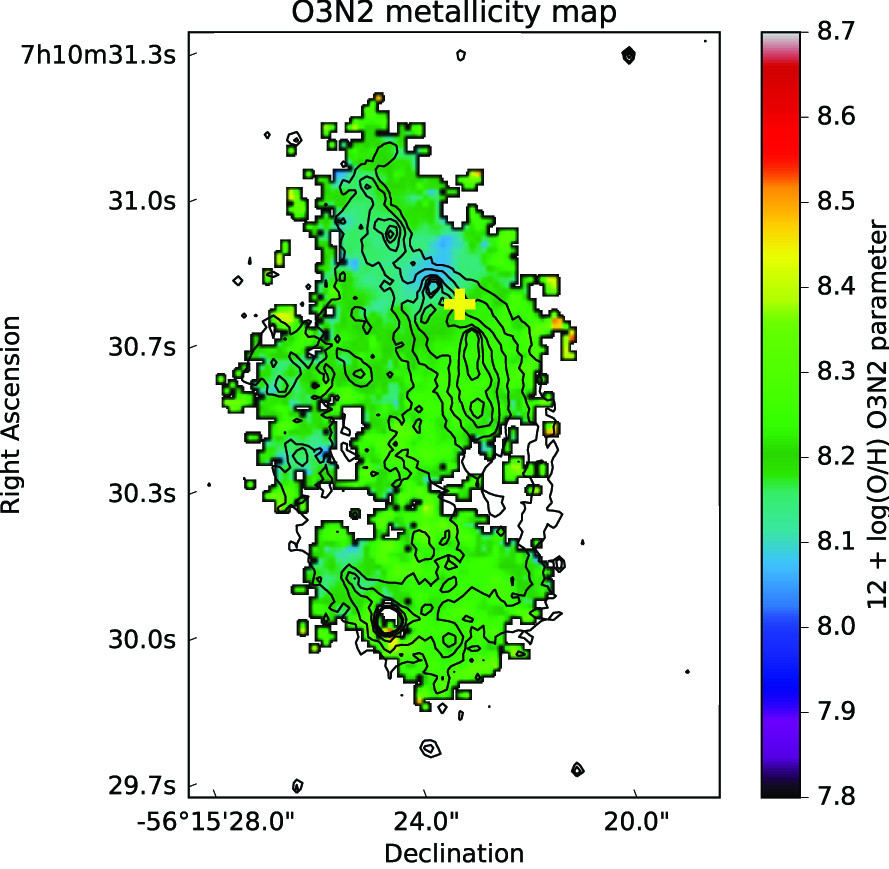}
	\includegraphics[width=0.65\columnwidth]{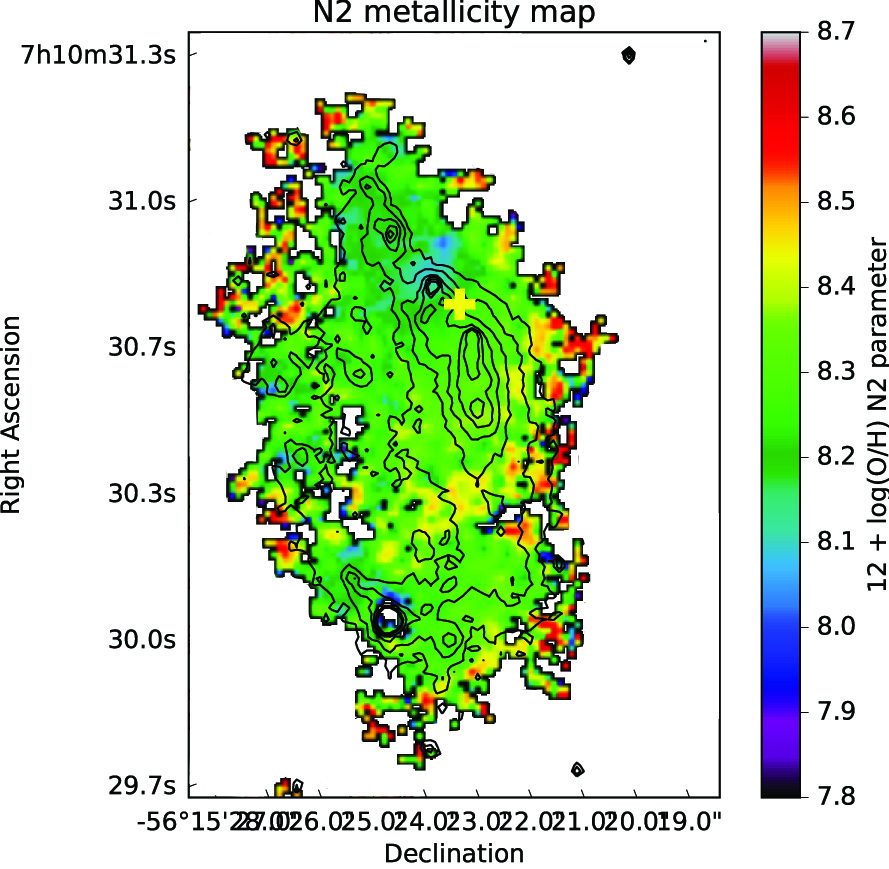}
	\includegraphics[width=0.65\columnwidth]{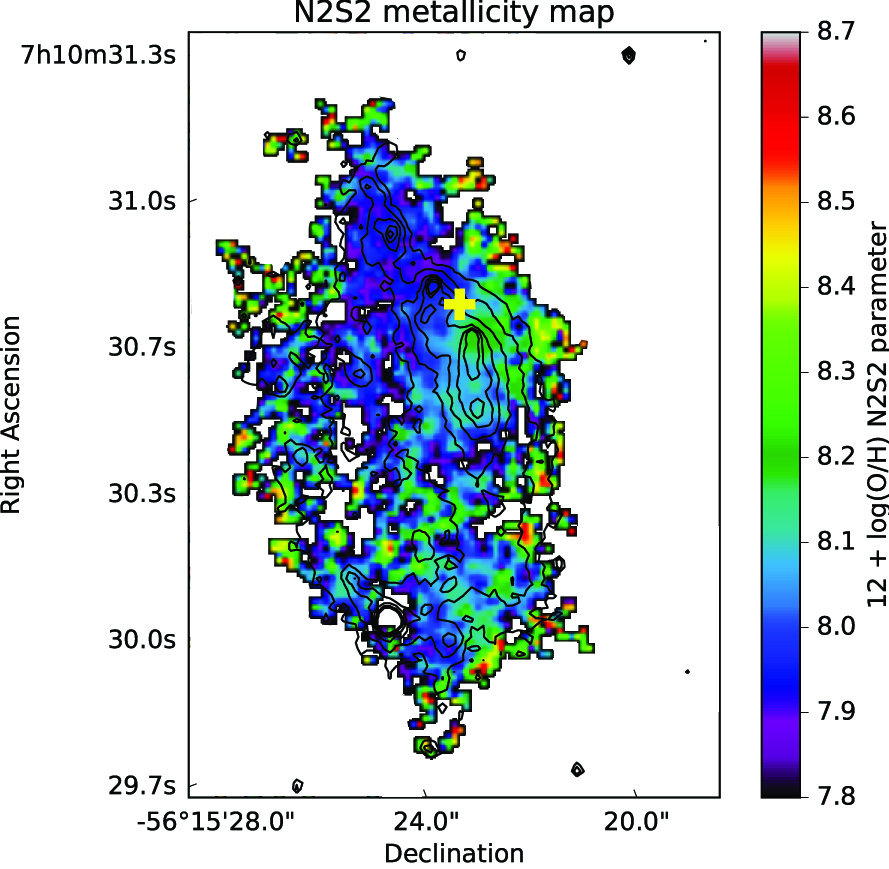}
    \caption{Metallicity maps obtained with the O3N2 (left), the N2 (center) and the N2S2 estimators using the formulation given in \citet{Marino2013} and in \citet{Dopita2016}. The GRB location is shown as a yellow cross.}
    \label{fig:no4}
\end{figure*}

We use two distinct metallicity estimators for the GRB site and the selected \ion{H}{II} regions, whose calibrations, given by \citet{Marino2013}, have been derived using objects with good determinations of electron temperatures using auroral lines: 1) the N2-index, whose analytical formulation is $12+log(O/H) = 8.743\,+\,0.462\, \times log([\ion{N}{II}] 6584/H\alpha)$, and 2) the O3N2-index, where $12+log(O/H) = 8.533\,--\,0.214\, log(([\ion{O}{III}] 5007/H\beta) \times (H\alpha/[\ion{N}{II}]6584))$. In Figure \ref{fig:no4} we show the corresponding metallicity maps while in Table \ref{tab:no2b} we report the values computed for the GRB and the \ion{H}{II} regions. Error measurements do not include the errors of the corresponding calibrators (0.16 dex and 0.18 dex for N2 and O3N2, respectively). We obtained similar values from both indices, although they are slightly higher for the N2 parameter with respect to the O3N2 (we report an average discrepancy of $\Delta$ = 0.04 between the N2 and O3N2 values). The GRB region is located in between two regions with the lowest (\ion{H}{II}-1) and the second highest (\ion{H}{II}-2) values for the O3N2 index. The highest metallicities, $12+log(O/H) = 8.25$ (O3N2), found for the \ion{H}{II}-5 and \ion{H}{II}-7 regions (8.10 in the \ion{H}{II}-2 region for the [\ion{N}{II}]/[\ion{S}{II}] index), imply that the entire galaxy shows a sub-solar metallicity. In the N2 map, we also note the presence of an extended area with a higher metallicity (log(O/H)+12 = 8.4-8.5), nearly perpendicular to the longer axis of the galaxy and close to its center.

Recently, \citet{Kruhler2017} reported in the analysis of IFU data of the host galaxy of GRB 980425 a strong dependence of the O3N2 and the N2 indices on the ionization parameters, leading to the conclusion that the observed variations of metallicity in single \ion{H}{II} regions are actually due to variations in the ionization, excluding possible metallicity gradients. Consequently they used an additional indicator for the metallicity, provided by the ratio of photo-ionized emission lines of H$\alpha$, [\ion{N}{II}] $\lambda$~6584 and the doublet [\ion{S}{II}] $\lambda\lambda$~6718/32 \citep[N2S2, ][]{Viironen2007,Dopita2016}. In the O3N2 and N2 maps, Fig. \ref{fig:no4}, of the GRB~100316D host, we note a metallicity gradient only for the \ion{H}{II}-1 region. We use also the N2S2 ratio formulation proposed by \citet{Dopita2016} as a metallicity estimator. The results are presented in Table \ref{tab:no2b}, while the spatial distribution of the N2S2 ratio is shown in the right panel of Fig. \ref{fig:no4}. Metallicities obtained with this method are generally lower than the O3N2 and N2 ones (but with larger uncertainties), in line with the results of \citet{Kruhler2017} who find higher values only for the Wolf-Rayet region detected in the GRB 980425 host galaxy. Inspection of the N2S2-based metallicity map reveals the same metallicity distribution observed also in the O3N2 and N2 maps: the region \ion{H}{II}-1 is characterized by the lowest metallicity observed in the galaxy, with the nearby GRB region showing a slightly higher value. 

An additional quantity that provides important constraints on the chemical evolution is the N/O ratio, which measures the relative abundance of nitrogen and oxygen \citep{Amorin2010,PerezMontero2014}. We used the freely available code HII-CHI-mistry\footnote{http://www.iaa.es/$\sim$epm/HII-CHI-mistry.html} which adopts a semi-empirical method, based on the comparison of observed emission line ratios with the predictions of a large grid of CLOUDY photoionisation models, to estimate the N/O value for the GRB and all the \ion{H}{II} regions. The code provides also an estimate of the 12 + log(O/H)$_{PM}$ value and of the ionization parameter (log $U_{PM}$) \citep{PerezMontero2014}. Results are shown in Tables \ref{tab:no2} and \ref{tab:no2b}. We do not observe a particular over-abundance of nitrogen in the GRB region as it is expected for Wolf-Rayet environments, but higher values are observed for the nearby \ion{H}{II}-2 region. In general, the GRB region is characterized by a low-metallicity ($Z_{GRB} = 0.3 Z_{\odot}$ if we consider the O3N2 and the N2 indices) but it is not the lowest value in the galaxy: the neighboring \ion{H}{II}-1 region shows the lowest values of metallicity in the entire galaxy. We will come back on this point later in the text.

\begin{figure}
	\includegraphics[width=\columnwidth]{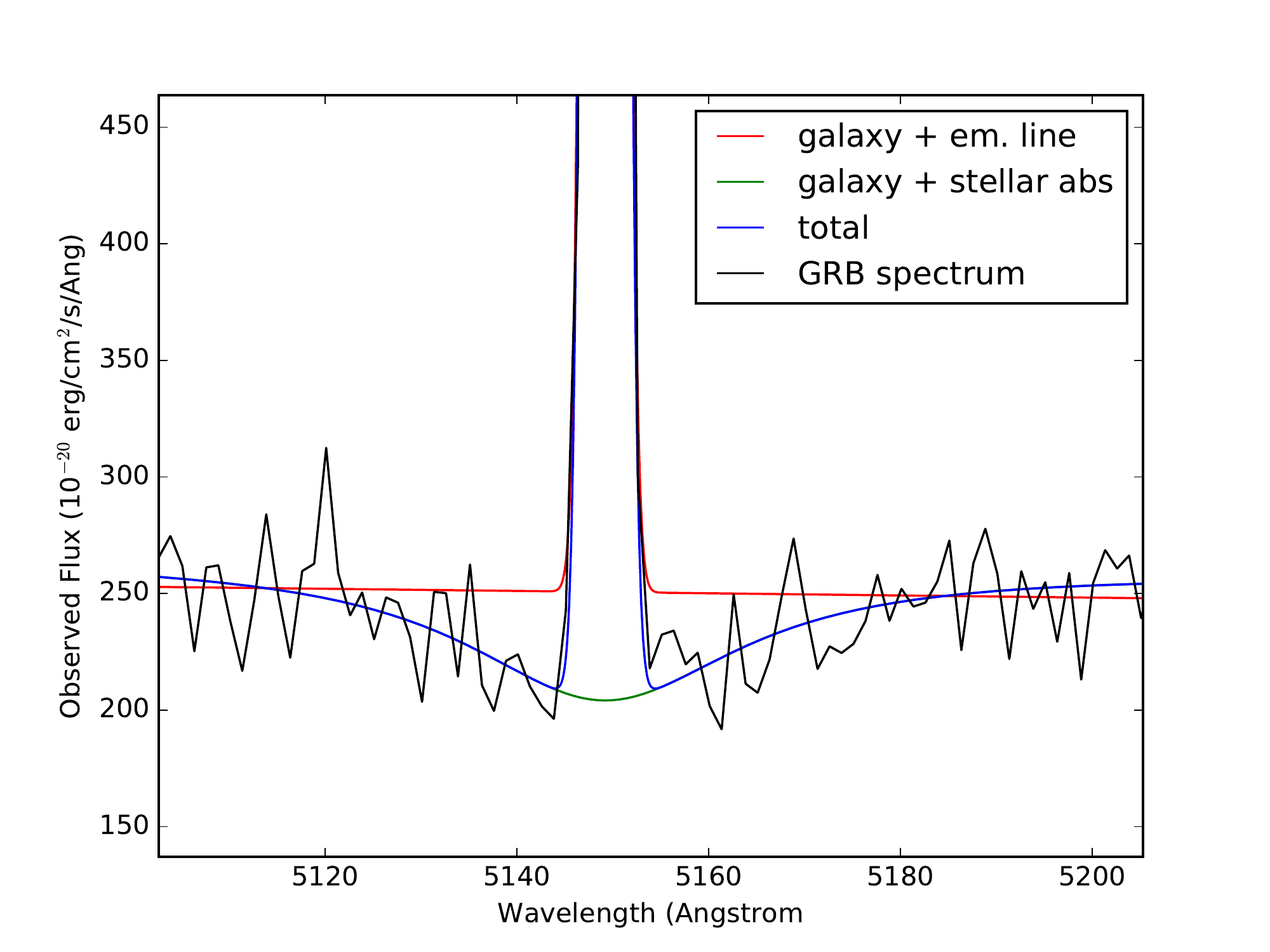}
    \caption{Stellar absorption component in the region around the H$\beta$ line for the GRB region, whose EW value has been used to estimate the stellar age. The models used to model the continuum, the stellar absorption and the emission lines are shown in different colors: 1) the galaxy continuum, modelled with a simple power-law function, and the emission line alone (red line); 2) the galaxy continuum, still with a power-law function, and the stellar absorption alone modelled with a Lorentzian profile (green line); 3) same as the case 2, but with the addition of a Gaussian function for the nebular emission line (blue line).}
    \label{fig:no3}
\end{figure}

\subsection{Ionization parameter}

The ionization parameter estimates the degree of ionization of the medium and is inversely related to the metallicity of the corresponding region \citep{Dopita2006}. It is an estimate of the ratio of the ionizing photon density  to  the  particle  density and can be estimated from the ratio of two lines of the same element that  correspond to different ionization states, such as the ratio between [\ion{O}{II}] $\lambda\lambda$~3727/29 and [\ion{O}{III}] $\lambda$~5007. In the MUSE data we do not observe the [\ion{O}{II}] $\lambda\lambda$~3727/29 doublet so we estimate the ionization parameter $\textrm{log\,U}$ with the following formulation \citep{Diaz2000}:

\begin{equation}
{\rm log\,U} = - 1.40\, {\rm log}([\ion{S}{II}]/H\beta) + 1.10\, {\rm log}(Z/Z_{\odot}) - 3.26,
\end{equation}
where [\ion{S}{II}] is the integrated flux of the ionized Sulphur doublet at 6718/32 $\lambda\lambda$, and log(Z/Z$_{\odot}$) = (12+log(O/H))$_{\textrm{obs}}$ - (12+log(O/H))$_{\odot}$ = (12+log(O/H))$_{\textrm{obs}}$ - 8.69 \citep{Asplund2009}. The GRB region shows a high ionization value of log\,U = -2.74 $\pm$ 0.01, see the second column in Table \ref{tab:no2}. The HII-CHI-mistry code also provides an estimate of the ionization parameter log U$_{\textrm{PM}}$. The results do not differ too much from the previous analysis, see third column in Table \ref{fig:no2}. In general, we report a low value for the ionization parameter with respect to other GRB host galaxies for which a measurement of log U has been reported \citep{Contini2017}.

\begin{figure}
	\includegraphics[width=\columnwidth]{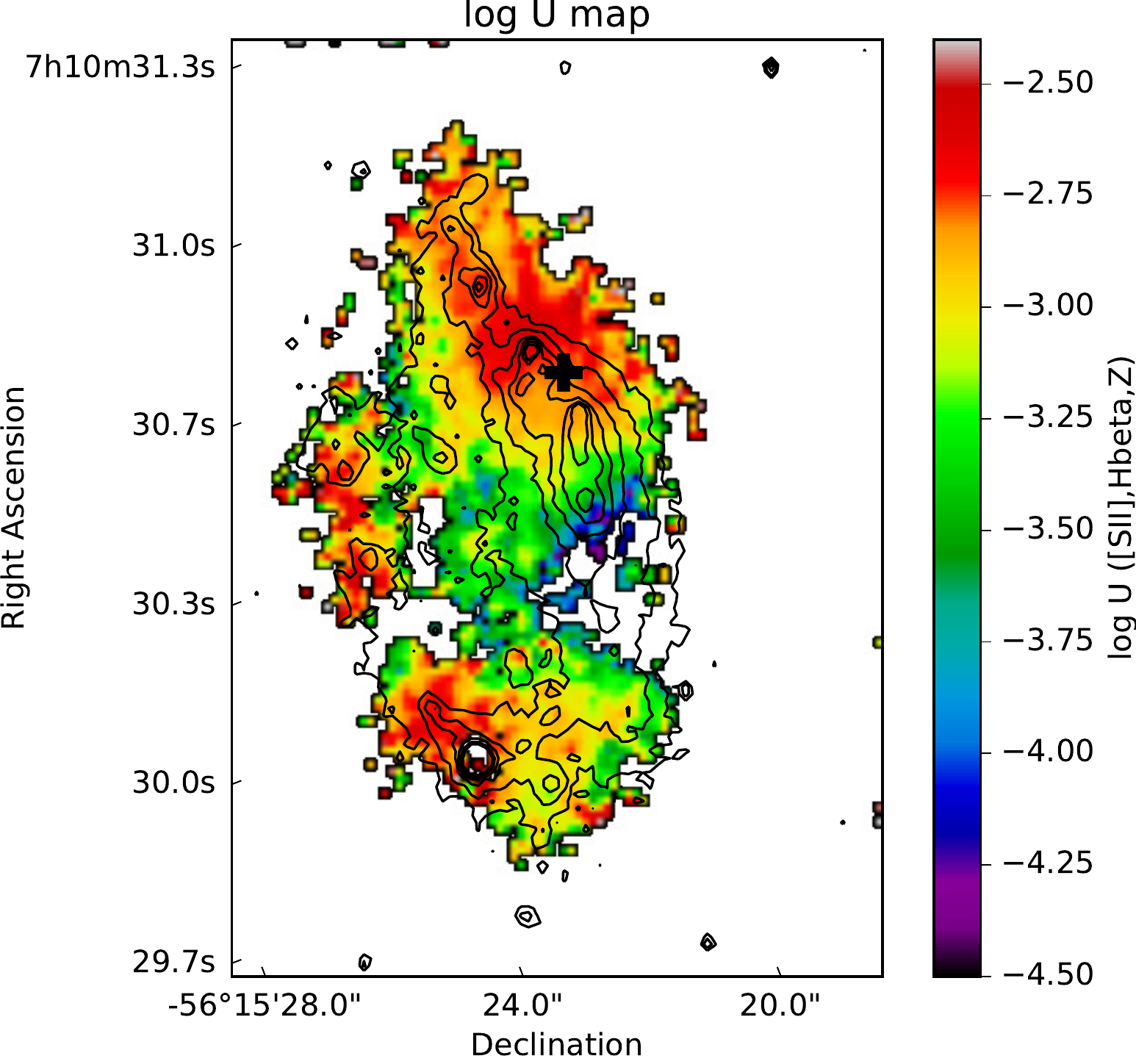}
    \caption{The ionization map obtained by using the [\ion{S}{II}] 6718/32 $\lambda\lambda$~estimator \citep{Diaz2000}. The GRB location is shown as a black cross.}
    \label{fig:no5}
\end{figure}

An interesting property worth checking is the presence of shocks in the galaxy, given its perturbed morphology. At optical wavelengths shocks are usually marked by the presence of strong emission lines from low-ionization species such as [\ion{N}{II]} $\lambda$ 6584, [\ion{S}{II}] $\lambda\lambda$ 6718/32 as well as [\ion{O}{I]} $\lambda$ 6300 and their ratio relative to H$\beta$ intensity. In addition, the signature of induced shocks is given by the presence of gas motions, generally tidally-induced flows \citep{MonrealIbero2010} or galactic winds \citep{Rich2010}, that can be identified by the presence of large velocity dispersion in the region where shocks are present, with respect to the rest of the galaxy \citep[see also][and references therein]{Rich2011}.

\begin{figure}
	\includegraphics[width=\columnwidth]{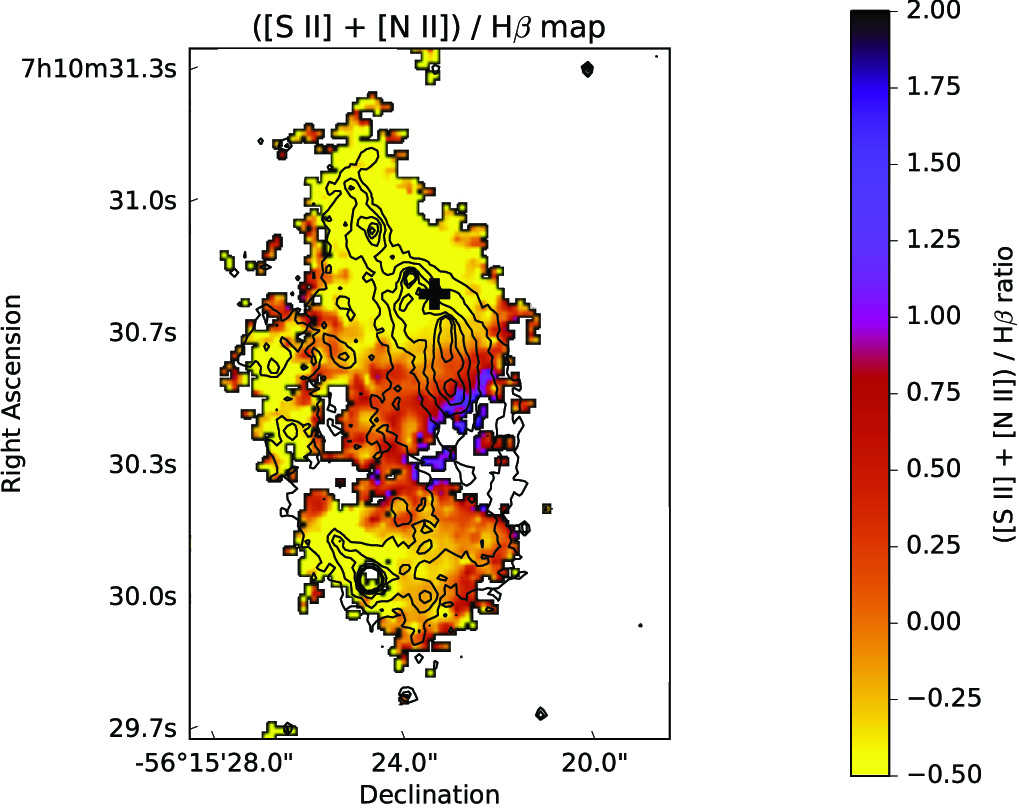}
    \caption{The distribution of the ([\ion{N}{II}] + [\ion{S}{II}]) / H$\beta$ in the host galaxy of GRB 100316D. Shocked regions are characterized by intense low-ionization emission lines like indeed [\ion{N}{II}] and [\ion{S}{II}] with respect to the H$\beta$ line: this map indicates in which regions of the galaxy there is a significant (large values of the index) presence of shocks due to interactions or internal perturbations.}
    \label{fig:app1}
\end{figure}

To this aim, we plot the galaxy distribution of the total flux of [\ion{N}{II}] and [\ion{S}{II}] with respect to H$\beta$, which is shown in Fig. \ref{fig:app1}. We find higher values far away from the \ion{H}{II} regions and close to the center of the galaxy. There is a gap in this central region of the galaxy, as it is visible in Figs. \ref{fig:no2a}, \ref{fig:no5} and \ref{fig:no6}, which is due to the cut-off in flux that we have chosen for single spaxels, in order to avoid noisy results in the maps. We also note that this central region shows the largest velocity dispersion inside the galaxy, as it is clearly visible in the right panel of Fig. \ref{fig:no8} (see also Section 5 for more details). To further investigate this region, we extracted an integrated spectrum of this specific region (with a size of 4 arcsec$^2$, see also Fig. \ref{fig:app2} in the Appendix), where we are able to measure the following emission line fluxes  (in units of 10$^{-18}$ erg/cm$^2$/s/\AA{}\,$^{-1}$), without having contamination by noise thanks to the sum over the spaxels:
H$\alpha$ = 27.7 $\pm$ 0.2, H$\beta$ = 14.1 $\pm$ 0.2, [\ion{O}{III}] $\lambda$ 5007 = 14.4 $\pm$ 0.2, [\ion{N}{II}] $\lambda$ 6584 = 5.6 $\pm$ 0.1, [\ion{S}{II}] $\lambda$ 6718 = 9.9 $\pm$ 0.1, [\ion{S}{II}] $\lambda$ 6732 = 6.9 $\pm$ 0.1. We indeed show in Fig. \ref{fig:app2} how the emission lines of low-ionization species, such as [\ion{N}{II}] $\lambda$ 6584 and [\ion{S}{II}] $\lambda\lambda$ 6717/32,  relative to H$\beta$, are stronger than the lines observed in the GRB region spectrum, which is a first evidence of the presence of shocks in this part of the galaxy.

\subsection{Star-Formation Rate}

\begin{figure*}
	\includegraphics[width=\columnwidth]{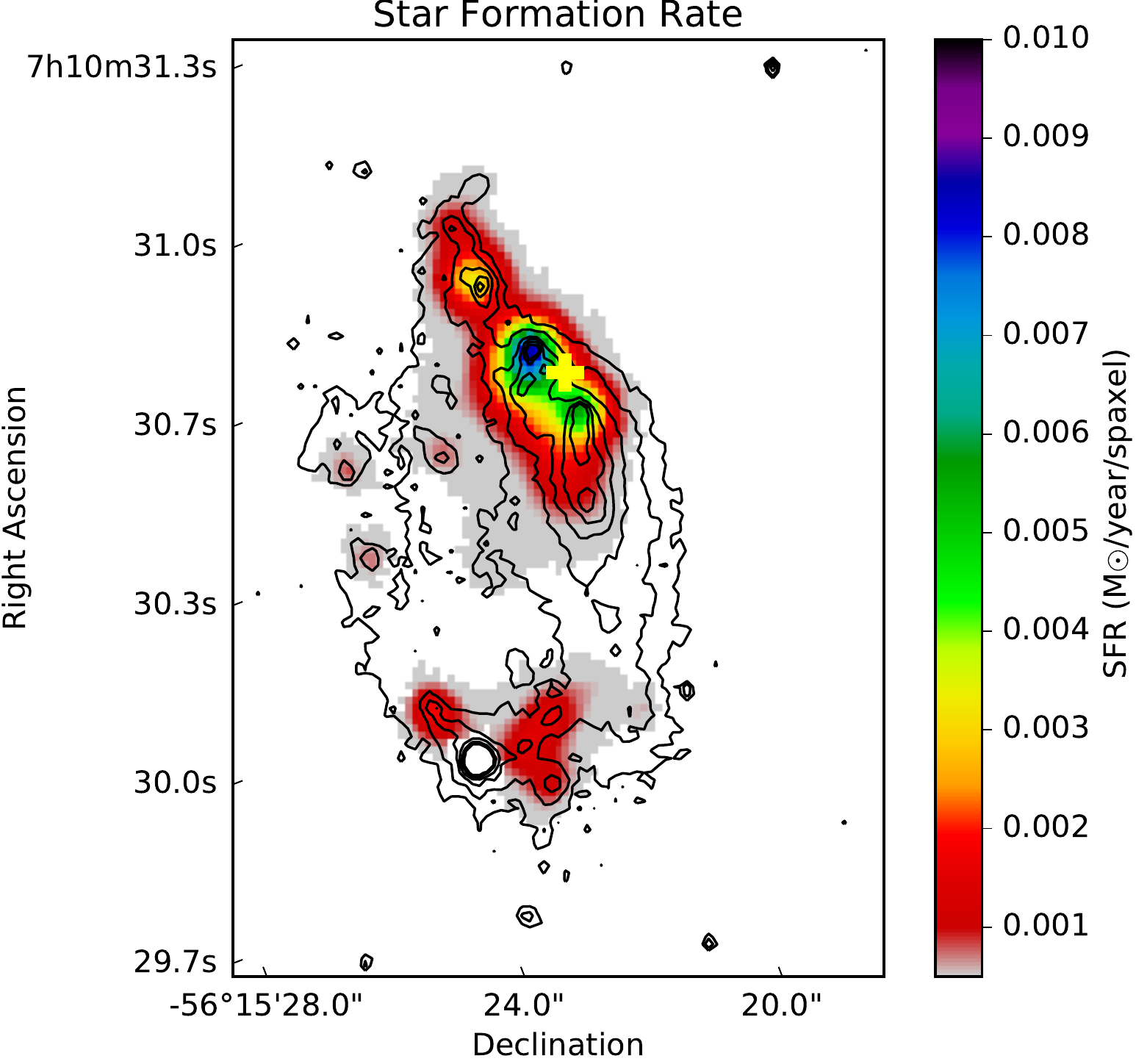}
    \includegraphics[width=0.96\columnwidth]{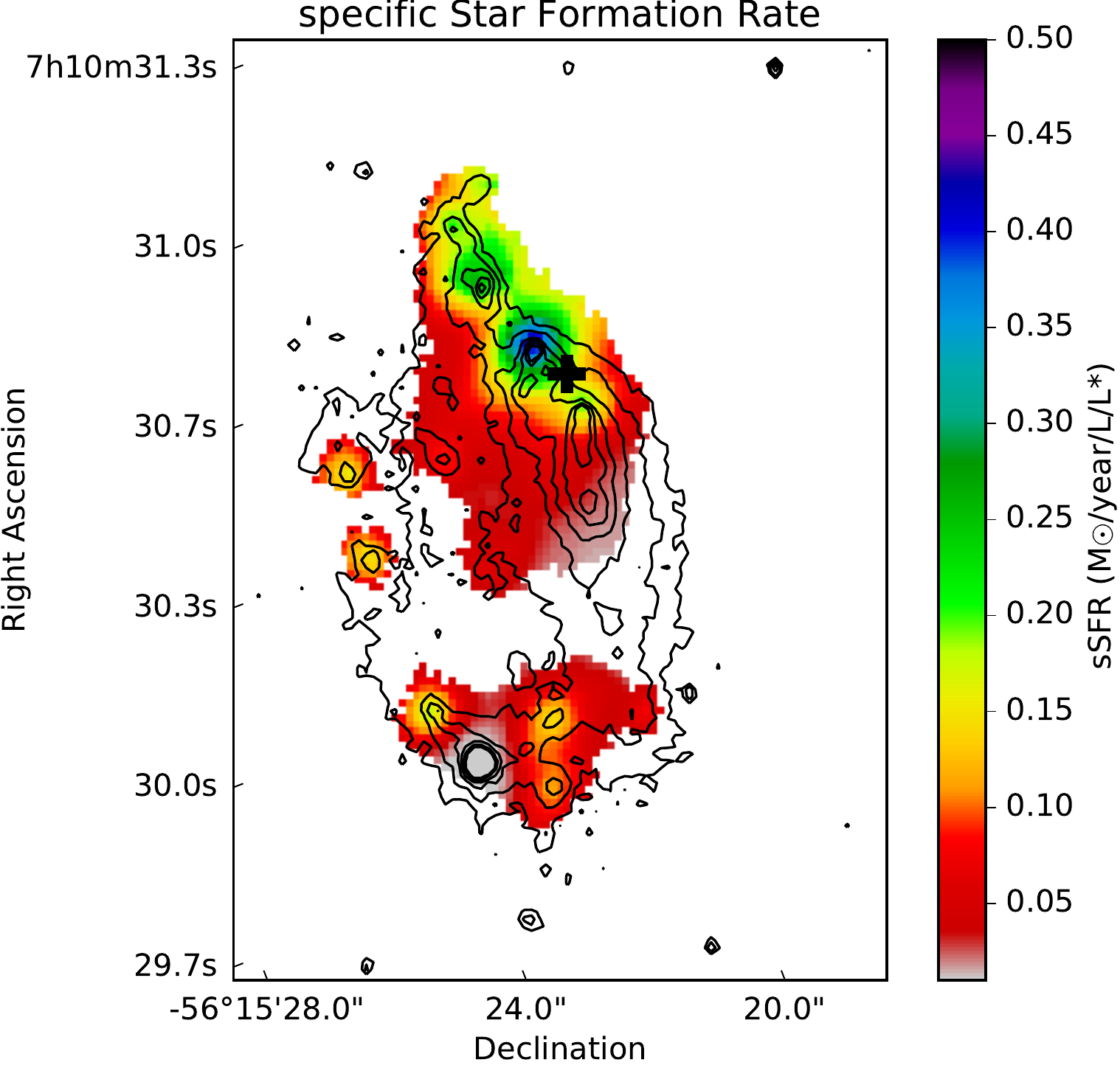}
    \caption{The left panel shows the SFR distribution in the GRB~100316D host galaxy  created through the H$\alpha$ flux diagnostic \citep{Kennicutt1989}, while the right panel shows the specific SFR obtained by weighting the SFR map and the rest-frame $B$-luminosity map, applying the same method described in \citet{Christensen2004}. The GRB location is shown as a yellow (left) and as a black cross (right), respectively. The area of one spaxel is 0.014 arcsec$^2$ (pixel dimension $0.117'' \times 0.117''$)}
    \label{fig:no6}
\end{figure*}

We determined the star-formation rate (SFR) using the H$\alpha$ line flux diagnostic applied to the flux-integrated and extinction-corrected region spectra as explained in \citet{Kennicutt1989}: SFR[M$_{\odot}$ yr$^{-1}$] = 7.9 $\times$ 10$^{-42}$ 4 $\pi$ d$_l^2$ F$_{H\alpha}$, where d$_l$ is the luminosity distance of the host galaxy. We find a star-formation rate in a region with a radius of $\sim$ 350 pc centered at the GRB location of SFR = 0.439 $\pm$ 0.001 M$_{\odot}$ yr$^{-1}$, see also Fig. \ref{fig:no6}, one of the highest values observed in the galaxy. The SFR shows large variations for different \ion{H}{II} regions: the complex system in the upper part of the galaxy, as visible in the left panel of Fig. \ref{fig:no6}, shows SFR values that are larger by an order of magnitude with respect to other \ion{H}{II} regions.

A comparison of current SFR with the averaged past one is provided by the luminosity-specific SFR, i.e. the star-formation rate per unit luminosity (sSFR = SFR/(L/L$^* $)). To compute this value we applied the method described in \citet{Christensen2004}, where a map of the rest-frame $B$-luminosity is weighted with the SFR map. We first derive a $B$-band rest-frame magnitude by extrapolating the spectrum for each region bluewards down to 4300 \AA\, after fitting the observed rest-frame part of the $B$-band continuum with a power-law function. We then integrate the obtained spectrum in the total rest-frame $B$-band filter transmission range. Finally, we estimate the luminosity by computing the corresponding absolute magnitude ($B = -18.4 \pm 0.6$ mag) and then multiplying the SFR map with the ratio between final absolute $B$-magnitude map and the magnitude of a M$_B$=-20.1 galaxy, an average value that has been inferred for the break in the Schechter luminosity function at redshift $z=0.04$ and $z=0.07$ from the analysis of blue galaxies in the DEEP2 and COMBO-17 surveys \citep{Faber2007}.

The final results for each \ion{H}{II} region are shown in Table \ref{tab:no2} and in Fig. \ref{fig:no6}, where it is evident that the GRB region is located close to the region with the highest specific star-forming activity, as expected for long GRBs \citep{Blanchard2016}. At the GRB site we find a luminosity-weighted specific SFR of sSFR = 0.38 $\pm$ 0.01 M$_{\odot}$yr$^{-1} / (L/L^*)$, which is similar to values obtained for other low-redshift GRB hosts \citep{Gorosabel2005,Sollerman2005,Christensen2008} and a bit higher than the values obtained for generic field galaxies \citep{Christensen2004}.

\subsection{Stellar age}

The age of the stellar-population at the GRB site is an indirect measure to constrain the mass of the progenitor star of the GRB. We  estimated the stellar age by using the underlying stellar absorption observed in the brightest Balmer lines, see Fig. \ref{fig:no3}, by using evolutionary stellar population synthesis codes, and assuming an initial Salpeter initial mass function (IMF) from 1 to 80 M$_{\odot}$. In the case of an instantaneous burst of star formation, it is possible to estimate the age of the underlying stellar population following the analysis presented in \citet{GonzalezDelgado1999}. To measure the EW of the stellar absorption, we fitted the data with a three-component model: 1) a power-law function for the galaxy continuum; 2) a Lorentzian profile function for the stellar absorption; 3) a Gaussian for the nebular emission line of the H$\beta$. With this model, we have measured an equivalent width of the H$\beta$ absorption for the GRB region (EW$_{H\beta_{abs}} = 5.7$\AA), that corresponds to an age of  20-30 Myr assuming a metallicity in between Z = 0.02 and Z = 0.001 (the model used does not provide estimates for Z = 0.004). We obtain similar values for the nearby \ion{H}{II} 1-2 regions, where the EW = 5.6-5.7 \AA. We also note that these values are in agreement and with the recent results obtained for the stellar population in the host galaxy of GRB 980425 associated with SN 1998bw, \citep{Kruhler2017}. 

An additional check of the previous result is provided by the nebular \ion{He}{I} lines: in the case of star-formation bursts older than 5 Myr, \ion{He}{I} lines are not observed in emission \citep{GonzalezDelgado1999}. When these lines are clearly detected, the ratio between the EW of \ion{He}{I} lines with respect to H$\beta$ provides an estimate for the stellar age. We clearly observe \ion{He}{I} $\lambda$~4922 in the spectrum of \ion{H}{II}-1 region, while it is marginally detected in the \ion{H}{II}-2 and the GRB region spectra. Following the results of \citet{GonzalezDelgado1999}, the ratio of \ion{He}{I} $\lambda$~4922 to H$\beta$ provides a younger stellar age of 5 $\pm$ 1 Myr for the \ion{H}{II}-1 region (EW$_{\ion{He}{I}}$ = 0.9 \AA\,), in the case of an instantaneous burst of star-formation. This result suggests that GRB~100316D was located in the region of the host with the youngest stellar age. In the following, given the absence of the \ion{He}{I} $\lambda$4922 in the GRB region spectrum, we will consider this value of 20-30 Myr as the age for the stellar population at the GRB site, although a younger value for the age cannot be completely ruled out.

Assuming instead a continuous star-formation rate with a constant value of 1 M$_{\odot}$ yr$^{-1}$ we obtain a stellar age larger than 30 Myr, based on the EW for Balmer lines and from \ion{He}{I} 4922 \AA\, line. 

\subsection{Electron density and temperature}

The radiation emitted in an \ion{H}{II} region depends on several factors such as the abundance, ionization, density and temperature of the gas. We have already provided an estimate of the ionization (see Sect. 3.3) as well as of the metallicity for each region (see Sect. 3.2). The electron density can be estimated by measuring the ratio between two lines of the same ion that are sensitive to the effects of collisional de-excitation \citep{Osterbrock}. The best estimate is provided by $p^3$ ions like [\ion{O}{II}] $\lambda\lambda$3727/29 and [\ion{S}{II}] $\lambda\lambda$6718/32. As already noted, we could not observe [\ion{O}{II}], so we relied on the [\ion{S}{II}] doublet. From the observed emission-line intensities, we estimated the local electron densities using an updated \citep{Proxauf2014} calibration of the density diagnostic described in \citet{Osterbrock}, obtaining an average density in the GRB region of $n_e \simeq 100$ cm$^{-3}$. The density at the GRB site is lower than typical values found in other GRB hosts by \citet{Savaglio2009}: in that work the authors used the [\ion{O}{II}] $\lambda\lambda$ 3727/28 doublet as the density estimator, obtaining density values ranging from 500 to 1300 cm$^{-3}$. In addition, extending the analysis to the entire host we note that the electron density distribution is higher at the GRB site than in the two adjacent \ion{H}{II} 1-2 regions, where it is particularly low (both have $n_{e} \eqsim 30$ cm$^{-3}$).

Similarly, the local temperature can be estimated from emission lines particularly sensitive to the temperature. This is the case of $p^2$ ions such as [\ion{O}{III}] and [\ion{N}{II}], where recombination lines arising from $^1$S $\rightarrow$ $^1$D and $^1$D $\rightarrow$ $^3$P provide a measurement of the local temperature, once we know the density. Unfortunately the [\ion{O}{III}] 4363 \AA\, line is not covered in the MUSE data and the other faint nebular line [\ion{N}{II}] 5755 \AA\ is visible only in the \ion{H}{II}-1 region spectrum with a flux of 0.17 $\times$ 10$^{-18}$ erg/cm$^2$/s. From the ratio \ion{N}{II} = (6550 + 6584)/(5756), and a density of $n_e \simeq 65$ cm$^{-3}$ for the \ion{H}{II}-1 region, we obtain a temperature value of $T = 12500 \pm 3000$ K.

\begin{table*}
	\centering
	\caption{Metallicity values, in 12 + log(O/H) units, derived for the GRB and all the \ion{H}{II} regions from three distinct indicators: the O3N2 and the N2 index as given in \citet{Marino2013}, the [\ion{N}{II}]/[\ion{S}{II}] ratio given by \citet{Dopita2016}, the semi-empirical approach explained in \citet{PerezMontero2014} and the N/O value provided by this latter method.}
	\label{tab:no2b}
\begin{tabular}{lccccc}
\hline
 Region   & O3N2          & N2            & N2S2  &  PM  & N/O  \\
\hline
 HII-1    & 8.14$\pm$0.0  & 8.16$\pm$0.01 & 7.95$\pm$0.01 & 8.38$\pm$0.01 & -1.47$\pm$0.01 \\
 HII-2    & 8.24$\pm$0.0  & 8.27$\pm$0.01 & 8.10$\pm$0.02  & 8.38$\pm$0.03 & -1.19$\pm$0.01 \\
 HII-3    & 8.16$\pm$0.01 & 8.18$\pm$0.01 & 7.92$\pm$0.02 & 8.34$\pm$0.02 & -1.45$\pm$0.01 \\
 HII-4    & 8.17$\pm$0.02 & 8.28$\pm$0.03 & 7.99$\pm$0.08 & 8.38$\pm$0.06 & -1.46$\pm$0.01 \\
 HII-5    & 8.25$\pm$0.01 & 8.27$\pm$0.01 & 8.04$\pm$0.03 & 8.52$\pm$0.02 & -1.40$\pm$0.01 \\
 HII-6    & 8.18$\pm$0.02 & 8.2$\pm$0.03  & 7.99$\pm$0.08 & 8.23$\pm$0.02 & -1.42$\pm$0.01 \\
 HII-7    & 8.25$\pm$0.01 & 8.18$\pm$0.02 & 7.89$\pm$0.05 & 8.41$\pm$0.08 & -1.46$\pm$0.01 \\
 HII-8    & 8.15$\pm$0.02 & 8.19$\pm$0.03 & 8.02$\pm$0.08 & 8.35$\pm$0.02 & -1.33$\pm$0.01 \\
 GRB      & 8.16$\pm$0.0  & 8.2$\pm$0.0   & 7.99$\pm$0.01 & 8.42$\pm$0.01 & -1.40$\pm$0.01 \\
\hline
\end{tabular}
\end{table*}

\subsection{Average properties of different H II regions}

\begin{figure*}
	\includegraphics[width=18cm]{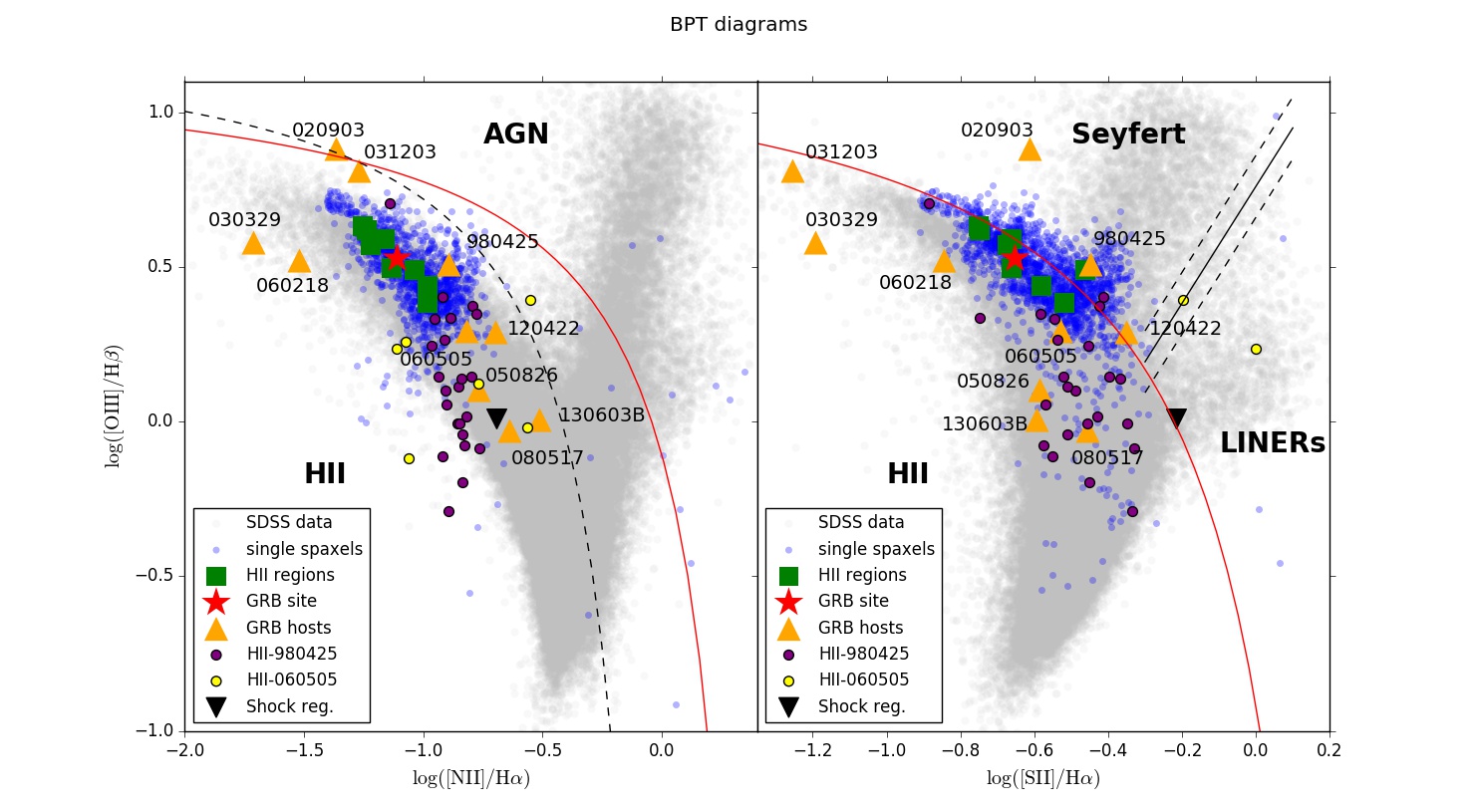}
    \caption{The BPT nebular emission-line diagnostic used to distinguish the ionization mechanism acting in \ion{H}{II} regions, and generally in the nebular gas in galaxies. There are two main versions of these diagrams:  the left side shows [\ion{O}{III}] 5007 \AA\, / H$\beta$ versus  [\ion{N}{II}] 6584 \AA\, / H$\alpha$, while the right sides substitutes [\ion{S}{II}] 6718/32 \AA\, doublet instead of [\ion{N}{II}]. The solid red and dashed black curves denote the divisions between star-forming dominated galaxies (below the lines) and active galaxies such as AGN, respectively, while in the right panel the upward-slanting black solid and dashed lines mark the division between active Seyfert galaxies and LINERs \citep{Kewley2001,Kauffman2003}. The single spectra for each spaxel of the host of GRB~100316D are shown as blue dots while integrated \ion{H}{II} region spectra are indicated as green squares. The black triangle represents the region where we find signature of shock interactions; this region, indeed, is located, although barely, in the LINER part of the BPT diagram. For comparison we also include the values obtained for integrated spectra of several GRB hosts (orange triangles) and individual \ion{H}{II} regions analyzed with IFU techniques in GRB 980425 \citep[violet circles, ][]{Christensen2008} and in GRB 060505 \citep[yellow circles, ][]{Thoene2014}. The grey data distribution corresponds to the SDSS-DR7 sample of galaxies described in \citet{Tremonti2004}.}
    \label{fig:no7}
\end{figure*}

\begin{figure}
	\includegraphics[width=1.1\columnwidth]{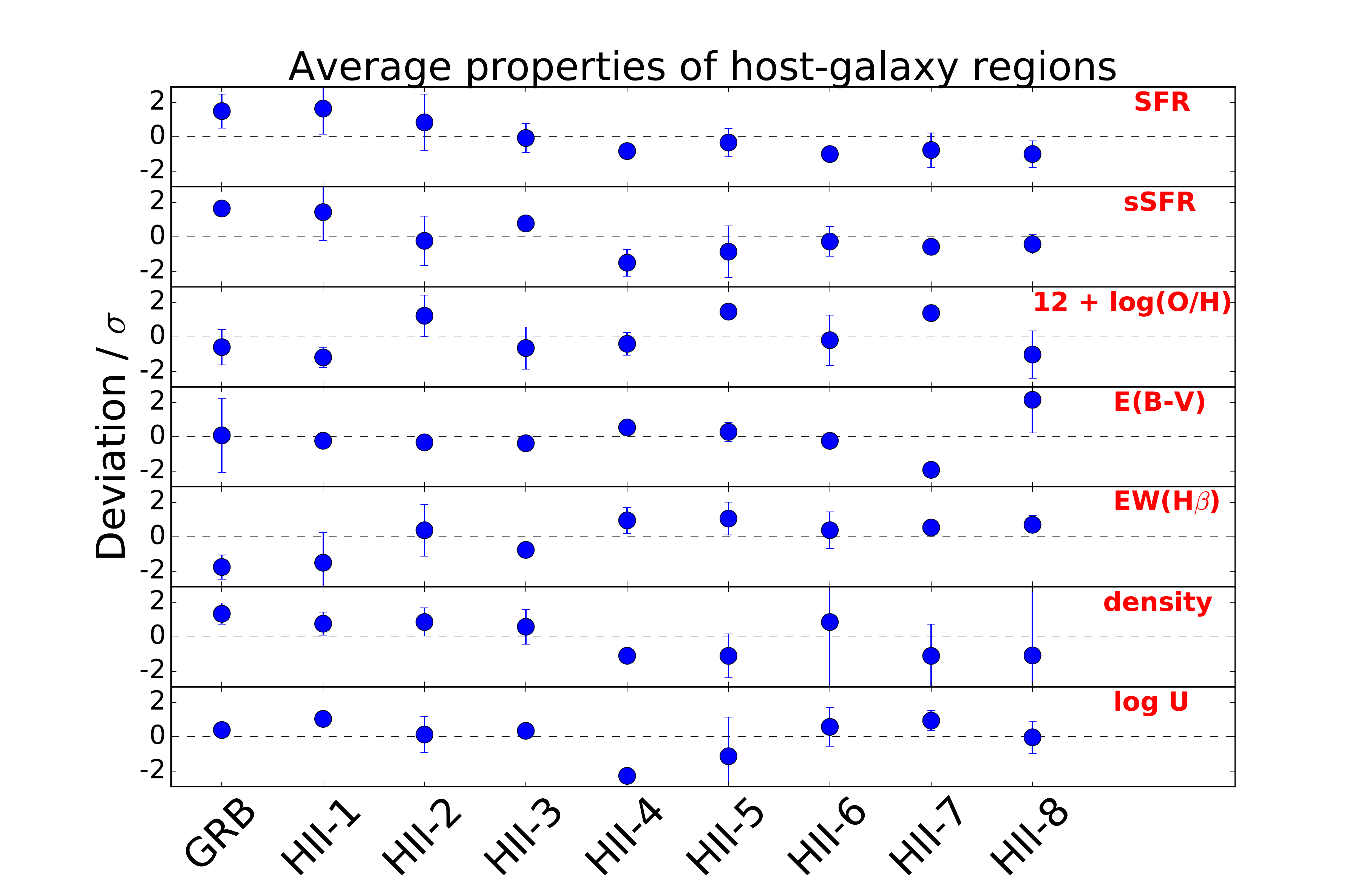}
    \caption{Deviation from the average value for each of the physical properties obtained for the \ion{H}{II} regions identified in the GRB~100316D host galaxy, as well as for the immediate GRB location.}
    \label{fig:no9}
\end{figure}

We investigate whether the GRB region shows different properties compared to the other \ion{H}{II} regions in the host. We showed before that the GRB location falls exactly in between the two brightest \ion{H}{II} regions of the galaxy. We compare here the results obtained for the GRB region with the properties of these two adjacent \ion{H}{II} regions as well as with the integrated properties of the entire host galaxy. 

We searched for the presence of active-galactic-nucleus (AGN) activity in the host galaxy applying the emission-line-ratio diagnostic for galaxy classification developed by \citet{Baldwin1981} (the BPT diagrams) which is used to identify the dominant excitation source of the region by radiation coming from young stars or excited by a Seyfert AGN or a LINER. This diagnostic tool is also useful to check for the presence of shocks in the nuclear region described in Section 3.3. The results of this analysis are shown in Fig. \ref{fig:no7} where the red line marks the division between objects characterized by \ion{H}{II} emission and by AGN activity \citep{Kewley2001}, respectively.

As initially reported in Section 3.3, emission-line ratios with LINER-like properties are also observed in case of shocks originating from galactic collisions or gravitational encounters \citep[see, e.g.,][and references therein]{Filippenko1996,Rich2011}: in these systems, a small amount of gas accreted from a small companion galaxy initiates cloud collisions that give rise to a LINER-like spectrum. We also expect to observe a broad H$\alpha$ line due to the velocity dispersion in this particular region. The region where we report the presence of shocks is marked as a black triangle in Fig. \ref{fig:no9} and is indeed located in the LINER region of the BPT diagram. We also include  emission-line ratios for other nearby ($z <$ 0.3) GRB host galaxies: GRB 980425 \citep{Christensen2008}, GRB 020903 \citep{Han2010}, GRB 030329 \citep{Levesque2010}, GRB 031203 \citep{Prochaska2004}, GRB 050826 \citep{Levesque2010}, GRB 060218 \citep{Levesque2010}, GRB 060505 \citep{Thoene2014}, GRB 080517 \citep{Stanway2015}, GRB 120422A \citep{Schulze2014} and the short GRB 130603B \citep{deUgarte2014}. Furthermore, we include the results obtained for individual \ion{H}{II} regions in GRB 980425 \citep{Christensen2008} and in GRB 060505 \citep{Thoene2014} analyzed by using IFU data. Although the progenitor of GRB 130603B was, very likely, the merger of two compact objects (being a short GRB) \citep{Tanvir2013}, we include it in this analysis to compare GRB/SN host galaxy properties with the ones inferred for the host of a completely different progenitor. 

The observed clustering in the BPT diagrams of \ion{H}{II} regions of the GRB 100316D host suggests that they share similar physical properties. We show in Fig. \ref{fig:no9} the variation of these physical properties by using the same analysis explained in \citet{Christensen2008} where it had been applied applied for the case of the host of GRB 980425. The GRB region shows a SFR value in line with the average value measured in the galaxy. The only large differences with the other regions consist in the low extinction value and in the measured high electron density, which is almost 3$\sigma$ larger than the average value obtained from the other \ion{H}{II} regions.

\subsection{The presence of [N I] $\lambda$5200 and forbidden iron lines}

\begin{figure}
	\includegraphics[width=\columnwidth]{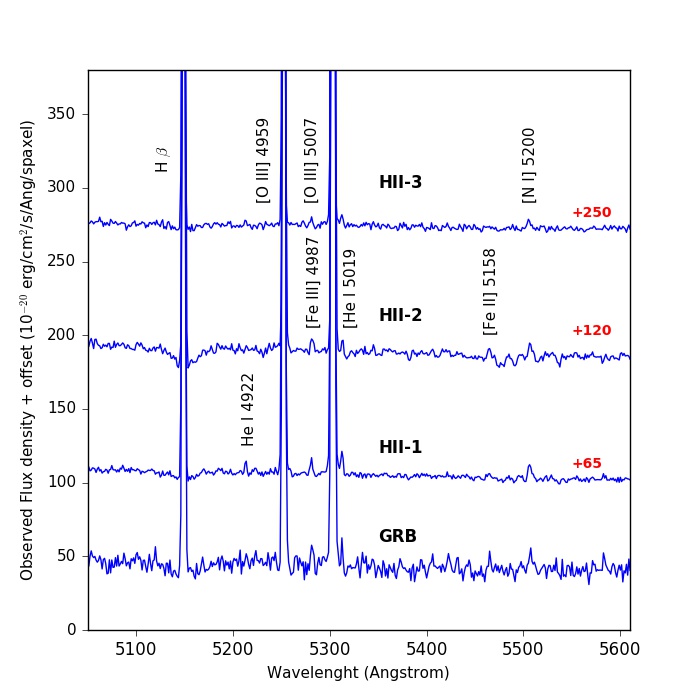}
    \caption{Selected wavelength region 5050-5600 \AA\, of the spectra of the \ion{H}{II}1-3 regions and the GRB location , showing the intensity of [\ion{N}{I}] 5200 \AA\,  with respect to  H$\beta$ . In the \ion{H}{II} 2 and GRB spectra, a weak [\ion{Fe}{II}] 5158 \AA\, line is also visible.}
    \label{fig:no9b}
\end{figure}

In the  brightest \ion{H}{II} regions of the host galaxy we detect the [\ion{N}{I}] 5200 \AA\, line while it is absent in the other ones. The spectrum of the immediate GRB region also shows a marginal detection of the line, see Fig. \ref{fig:no9b}. This line, actually a doublet separated by only 2 \AA\,, is usually observed to be as bright as H$\beta$ in planetary nebulae, in supernova remnants such as the Crab nebula and in the core of galaxy clusters \citep{Ferland2009}, as well as in nearby \ion{H}{II} regions like the Orion nebula \citep{Baldwin2000}. In this latter case, \citet{Ferland2012} have shown that the [\ion{N}{I}] 5200 \AA\, line is produced by pumping via far-ultraviolet stellar radiation at wavelengths near the \ion{N}{I} lines (951 - 1161 \AA\,) with an important contribution from an underlying non-thermal component that drives the process. In the specific case of the Orion nebula, the ratio between the fluxes of [\ion{N}{I}] 5200 \AA\, and H$\beta$ ranges between the interval $I([\ion{N}{I}])/I(H\beta )$ = 10$^{-3}$--10$^{-2}$, and is larger for the companion M43 nebula. This discrepancy is likely attributed to the different spectral types of the underlying illuminating stars (O7 type for the Orion nebula, B0.5 for the De Mairan's Nebula, M43); indeed \citet{Ferland2012} found that the above flux ratio scales with the ratio between the far ultraviolet and the extreme ultraviolet radiation from the illuminating stars. For the \ion{H}{II}-1,2 and the GRB~100316D regions we find values of $I([\ion{N}{I}])/I(H\beta ) = 0.02-0.03$, which suggest a population of underlying stars of late O -- early B spectral types and thus confirm the young age inferred for the stellar population in these regions. Assuming these stars formed during an instantaneous burst of SF $\sim$ 20-30 Myr ago, it is possible to detect additional SN or GRB explosions from this region in the next 5-10 Myrs.

In the vicinity of the [\ion{N}{I}] line we note a weak emission line observed in the spectrum of the \ion{H}{II}-2 and the GRB regions, which we attribute to the low-ionized [\ion{Fe}{II}] 5158 \AA\, line, see Fig. \ref{fig:no9b}. The combined presence of this line with the [\ion{Fe}{III}]~$\lambda\lambda$~4659,4987 lines in the same spectrum suggests that iron in these spectra is very likely collisionally excited by thermal electrons and then cools in denser environments. A shock originating from the passage of a SN blast-wave or from internal motion in the \ion{H}{II} region is the most plausible cause. Deeper spectroscopic observations can provide better measurements of the physical properties (shocks parameters, local density, temperature) with the use of numerical simulations. 

\section{The host galaxy of GRB 100316D at higher redshifts}
As demonstrated in the previous sections, MUSE has great potential to spatially resolve the host galaxies of GRBs at low redshift in one shot without slit losses. However, the peak of the redshift distribution of GRBs is at $z=2.1-2.2$ \citep{Jakobsson2012,Perley2016} and the most used technique to investigate GRB host galaxies at high redshifts consists of using long-slit spectroscopy. The information obtained from this analysis can however be biased, depending on the specific integrated region of the host galaxy that enters the slit, with the important consequence that we do not have information on the immediate environment of the GRB. In addition, the physical properties measured through emission lines in long-slit spectra also provide a picture of the total galaxy, instead of measuring the immediate GRB environment.

For this reason we simulated a long-slit spectrum of the host galaxy of GRB 100316D by integrating the spaxels inside the area corresponding to a slit for the theoretical case of the host being at redshift $z=2$, which includes the entire host galaxy. However, we exclude the bright star near the \ion{H}{II}-4 and \ion{H}{II}-5 regions, to avoid possible contaminations from this source. We also apply a correction to the observed flux  for the distance. At $z=2$ several emission lines useful for the computation of the ionisation parameter and metallicity would be redshifted into a range difficult to observe with ground-based telescopes, but other lines such as the [\ion{O}{II}] $\lambda\lambda$~3727/29 doublet would be visible, so we discard any observational problem for the following analysis (although in this case we could measure metallicity with the direct method). Finally, we measured all the physical properties in this simulated spectrum and compared them with the same quantities observed for the GRB region, see Table \ref{tab:no4}.

\begin{table}
	\centering
	\caption{Comparison between the physical properties measured for the GRB region and the total integrated host galaxy assuming it would be observed at redshift $z=2$.}
	\label{tab:no4}
\begin{tabular}{lcc}
\hline
 Physical   & Host galaxy          &  GRB \\
 Indicator  & at $z = 2$         &  region  \\
\hline
O3N2    &  8.18$\pm$0.16 & 8.16$\pm$0.01 \\
N2   & 8.25$\pm$0.24  & 8.21$\pm$0.01  \\
N2S2   & 7.99$\pm$0.63 & 8.00$\pm$0.01  \\
log U   & -2.98$\pm$0.58 & -2.76$\pm$0.01  \\
E(B-V)   & 0.57$\pm$0.14 & 0.46$\pm$0.03  \\
SFR  & 1.20$\pm$0.08 & 0.439$\pm$0.001  \\
EW(H$\beta$)  & -18.7$\pm$5.4   & -75.3$\pm$18.9   \\
EW(H$\alpha$)  & -87.6$\pm$11.7  & -229.2$\pm$18.9 \\
\hline
\end{tabular}
\end{table}

We note a good agreement for the three distinct metallicity indicators used, while for the other properties several significant differences are revealed: the ionisation parameter value obtained from the simulated long-slit spectrum is lower than the measured log\,U for all \ion{H}{II} regions as well as for the GRB spectrum. Moreover, a large difference is found for the extinction:  we have already reported above that the GRB region shows the lowest extinction value in the entire galaxy, while the integrated one is twice the GRB value. Finally, the EW values measured for the Balmer lines in the simulated spectrum are lower than the ones observed for many (if not all, as for the case of the H$\beta$ line) \ion{H}{II} regions, an effect attributed to the integration over the total area of the galaxy which includes regions that are affected by the host-galaxy continuum. 

\section{The search for possible companions of the GRB 100316D host}

\begin{figure}
	\includegraphics[width=\columnwidth]{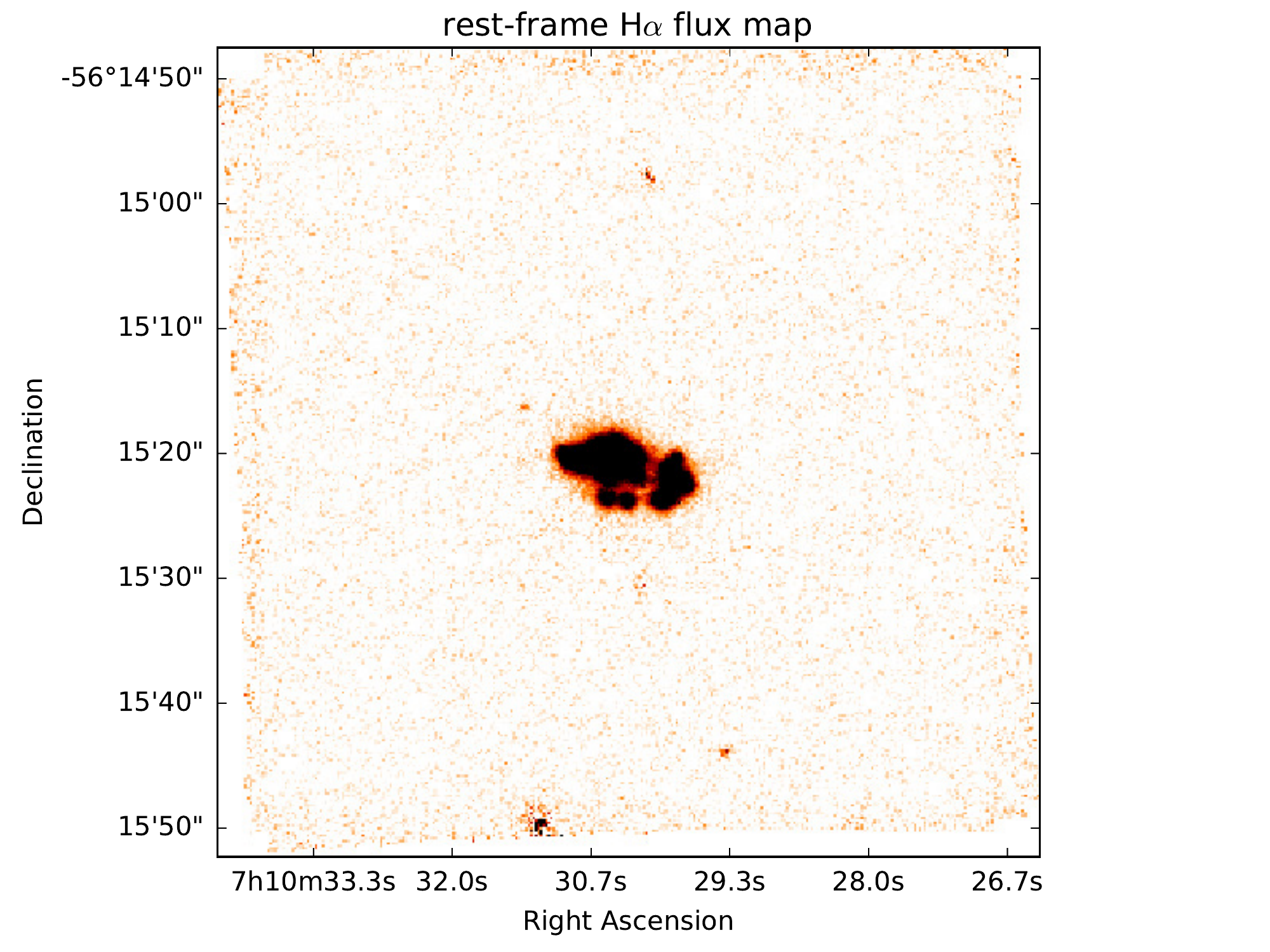}
    \caption{The rest-frame H$\alpha$ flux map obtained for of the entire MUSE field of view. The GRB 100316D host galaxy is at the center of the image. We do not detect any galaxy companions of the host up to a flux in $H\alpha$ of $2 \times 10^{-19}$ erg/cm$^2$/s, which corresponds to a SFR upper limit of $SFR_{c} \leq 10^{-5}$ M$_{\odot}$ yr$^{-1}$. North is upward while East is leftward.}
    \label{fig:no8b}
\end{figure}

\begin{figure*}
	\includegraphics[width=\columnwidth]{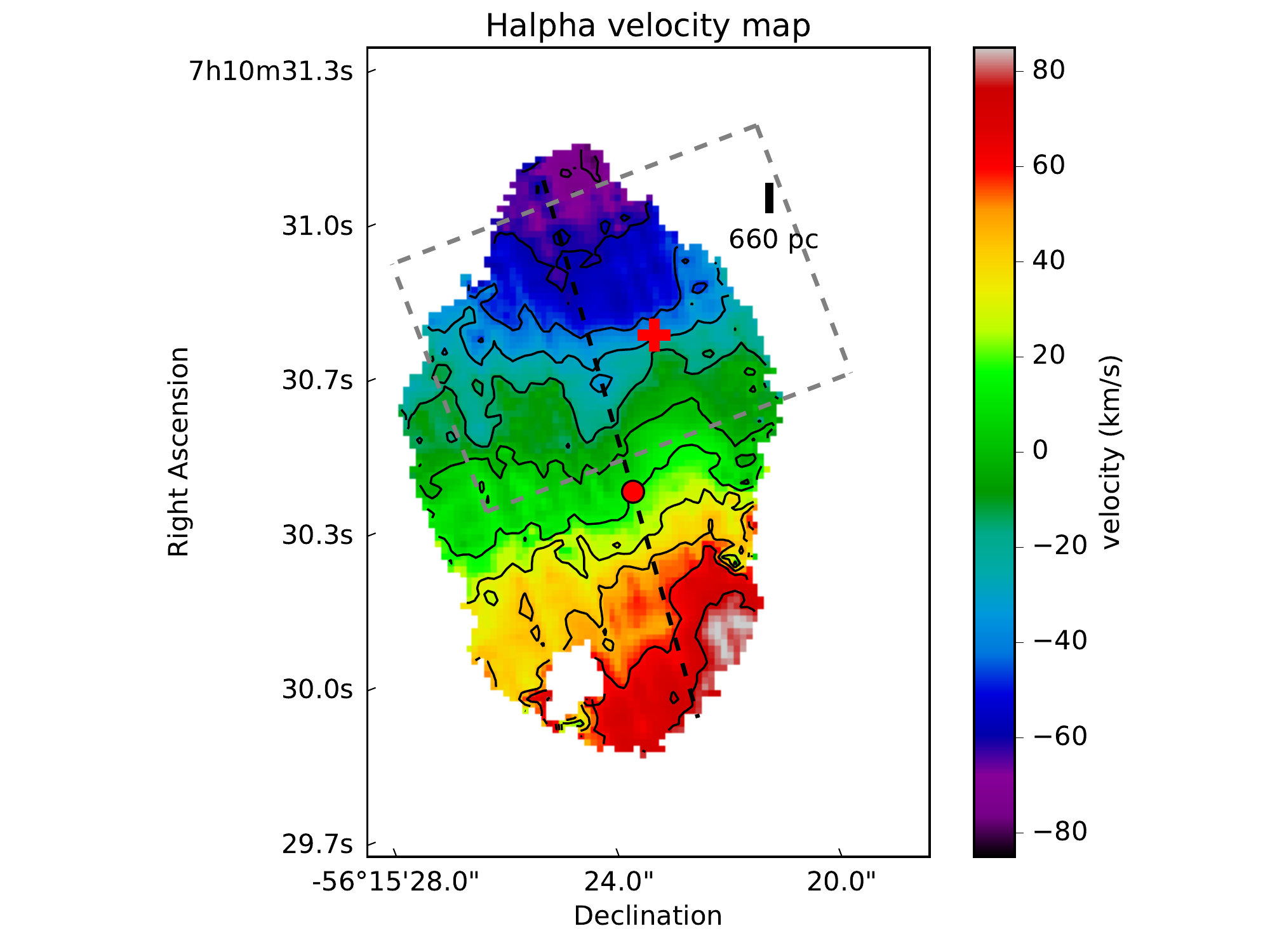}
        \includegraphics[width=\columnwidth]{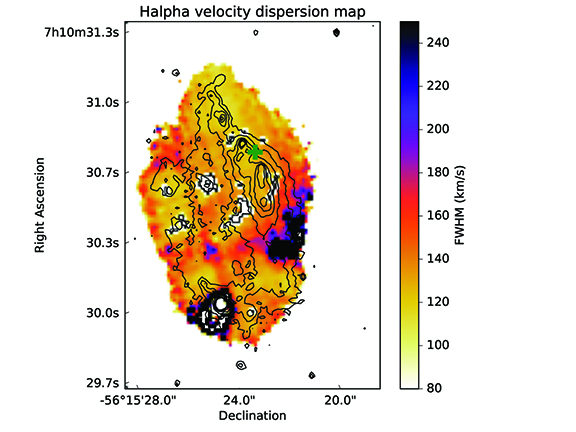}
    \caption{\textit{(Left panel)} The velocity distribution obtained from fitting the H$\alpha$ line throughout the entire GRB~100316D host galaxy. Iso-velocity contours spaced at 15 km/s from each other are shown as black curves. The zero-velocity location is reported as a red circle. The dashed gray rectangle corresponds to the  FLAMES field-of-view. The black dashed line corresponds to the kinematic major axis. The distance scale projected on a 65 degree axis is reported in the upper right. \textit{(Right panel)} The distribution of the FWHM obtained from fitting the H$\alpha$ line in each spaxel over the entire GRB~100316D host galaxy. The GRB location is shown as a red and a green cross, respectively. Large values for the dispersion velocity correspond to the galaxy region where we observe possible shock-signatures.}
    \label{fig:no8}
\end{figure*}

Its luminosity ($M_B = -18.4$ mag) and size ($\sim 12$ kpc) suggest that the host galaxy of GRB~100316D could be an irregular dwarf galaxy. Morphologically, these galaxies are different from ''normal'' spiral galaxies: they do not show a clear bulge in their center or only a very small bulge component whose main contribution is important in the early rising part of the rotation curves. However, the observed disturbed morphology, see Fig. \ref{fig:no1}, does not exclude that the entire system is interacting and the active star-bursting phase is due to an on-going recent merger with some small nearby companions. It is possible to test this hypothesis studying the velocity map of the galaxy in order to identify anomalies and/or inhomogeneities in the velocity field, as well as investigating in the entire field of view of MUSE for the presence of interacting companions. In Fig. \ref{fig:no8b} the rest-frame H$\alpha$ map obtained for the total MUSE field of view is shown. We do not detect the presence of companions around the GRB 100316D host galaxy: we estimate a three sigma observed flux upper limit, at the red-shifted H$\alpha$ wavelength, of $2 \times 10^{-19}$ erg/cm$^2$/s, which corresponds to an upper limit in luminosity of $L_{H\alpha} \leq 10^{36}$ erg/s. The small point-like concentrations of H$\alpha$  in the upper-left and just beneath the bright host galaxy complex are due to noisy residuals by foreground stars.

\section{Kinematics}

We can study the velocity distribution across the entire host galaxy of GRB~100316D by measuring the shift of the observed bright emission lines with respect to the corresponding zero velocity wavelength. To determine this latter value, we first fit the H$\alpha$ and the [\ion{O}{III}] 5007 \AA\, lines in the different spaxels by using an automatic Gaussian-line fitting procedure, obtaining as zero velocity wavelength value $\lambda_{0} = 6954.42\, \AA\,$, corresponding to the redshift $z = 0.05907$, which we consider as our reference value. In Fig. \ref{fig:no8} we present the iso-velocity contour map obtained using the H$\alpha$ line where we see a homogeneous velocity distribution along the East-West direction in both maps, which suggests an almost uniform rotation of the galaxy around a common center with some deviations that correspond to observed large \ion{H}{II} regions (see Fig. \ref{fig:no8}). We also note the presence of a central region characterized by large FWHM values (right panel in Fig. \ref{fig:no8}): this region corresponds to the area of the galaxy where we have found shock signatures and a LINER-like spectrum.

In order to extract a rotation curve, we need to first  determine the kinematic major axis. Since the center of the galaxy is not clearly identified in the galaxy image, we first determine a rough orientation for the kinematic major axis of the galaxy from the velocity field and then build a rotation curve by iterating and extracting several possible kinematic axes and finally averaging all the obtained curves. The center of the galaxy is then identified as the location with a radial velocity equal to zero. The host inclination angle has been estimated from the ratio of the major to the the minor kinematic axes, as considered from the center: $i = 65 \pm 5$ deg. 
We then extract the rotation curve along the kinematic major axis. Since the spectral resolution of the FLAMES dataset is better than that of MUSE, we have also used the FLAMES dataset for the fit of the rotation curve. After correcting the FLAMES cube for the astrometry, we note that FLAMES data do not cover the entire host galaxy, see left panel in Fig. \ref{fig:no8}, and the final rotation curve refers only to the blue-shifted side of the galaxy. The rotation curve is very detailed, with a good spatial sampling (at the redshift of the host, we have 1.0$''$ = 1.15 kpc) and resembles the typical rotation curves of dwarf galaxies \citep[characterized by a uniform slowly rising behavior,][]{SparkeGallagher2007}, evidence that is also sustained by the stellar mass and the absolute magnitude of the host: from the rest-frame $B$ filter map (obtained by extrapolating the derived host galaxy SED to $\lambda$~=~4215~\AA) we estimate an absolute integrated magnitude of $M_B = -18.4 \pm 0.6$ mag. 

\begin{figure}
	\includegraphics[width=1.1\columnwidth]{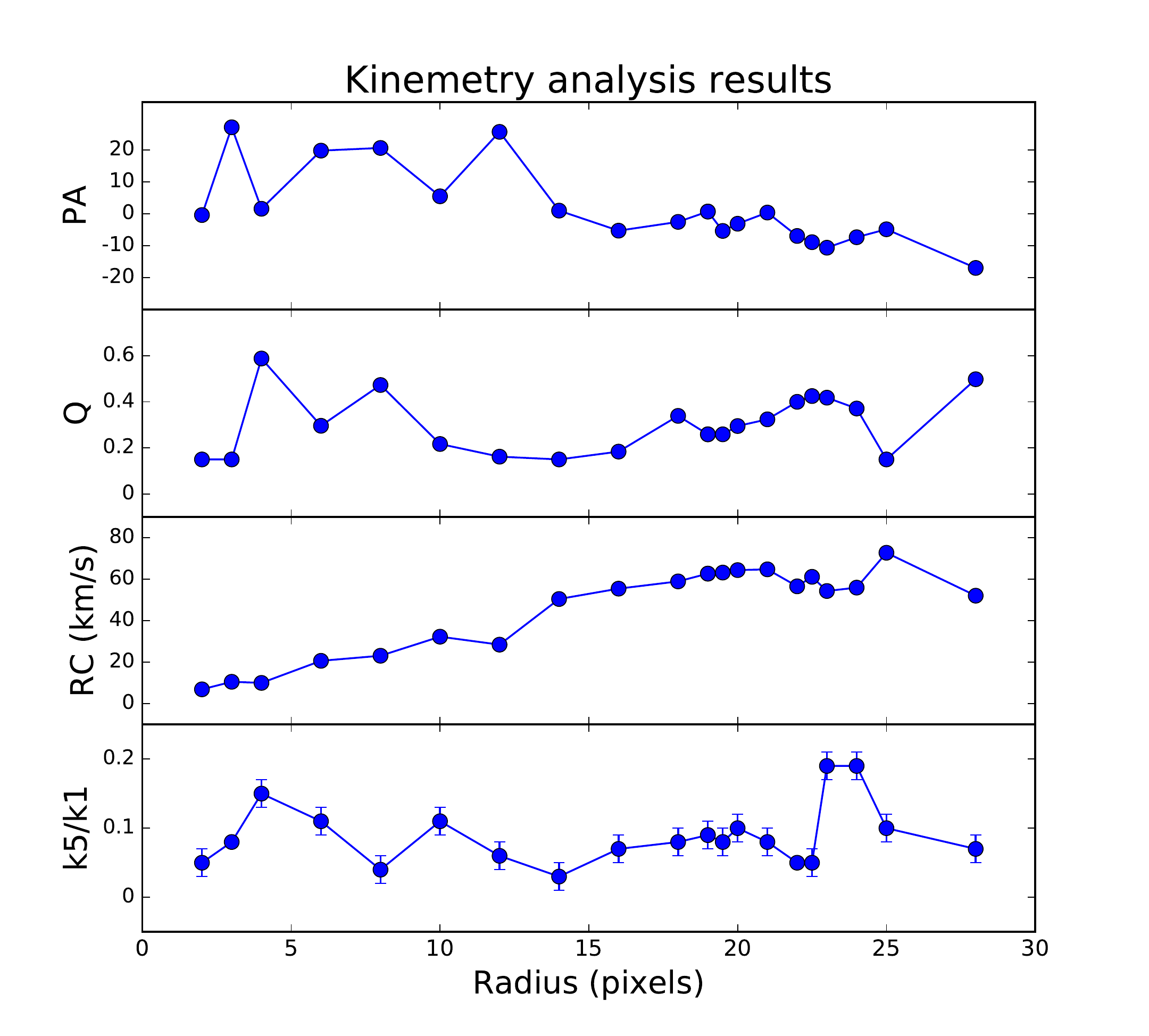}
    \caption{Best-fit parameters obtained from the kinemetry analysis \citep{Krajnovic2006} for different positions (1 pixel corresponds to 0.2$''$) of the tilted-rings relative to the center of the GRB 100316D host galaxy. From the top: Position angle in degrees, the ellipticity parameter Q,  the rotation curve (RC) that corresponds to the first order $k1$}, and  the $k5/k1$ ratio. The distribution along the entire galaxy of this last parameter suggests that the entire system is out of equilibrium. Note also the similarity of the rotation curve with the one extracted from the kinematic major axis (see Fig. \ref{fig:no10}). 
   \label{fig:no9bb}
\end{figure}

We also performed a higher-order analysis of the line-of-sight velocity distribution with kinemetry models \citep{Krajnovic2006} that have been successfully applied to several galaxies \citep{Krajnovic2008} and to the host of GRB 060505 \citep{Thoene2014}. This analysis consists of dividing the velocity map into multiple tilted-rings and then performing an harmonic expansion analysis along these rings in order to extract additional kinematic information, such as the presence of additional kinematic components, the velocity field and the presence of perturbations. One of the main requisites of the method is that the emission-line velocity indicators must be confined to a thin disc structure. The most important Fourier expansion terms are $k1$, which represents the velocity amplitude, and the first higher order that is not fitted, the $k5$ term, usually normalized with respect to $k1$. The presence of induced shock interactions in the GRB 100316D host would be evidence of a system that is not completely virialized. Indeed, due to their low gravitational energy, with respect to large and massive spirals, a gravitational encounter of a low-mass galaxy with a small and faint companion is able to affect strongly its dynamics. One of the main observational evidences is then given by the presence of multiple kinematic components, as derived from the analysis of its rotation curve.
Indeed, the result of the kinemetry analysis shows a non-viralized system, as indicated by the $k5/k1$ ratio in Fig. \ref{fig:no9bb}. The presence of multiple peaks characterized\footnote{The limiting value that divides single component disc-like rotations from multiple kinematical components depends on the IFU used. In \cite{Krajnovic2008} the conservative value of $k5/k1 = 0.04$ is considered.} by $k5/k1 \approx 0.2$ points to the existence of additional kinematic components/perturbations distributed along the entire galaxy, in particular in the external regions (the peak in the $k5/k1$ plot around radius values of 22-25 pixels in Fig. \ref{fig:no9bb}) where the large star-forming regions and that one of the GRB are located. These perturbations reflect a turbulent internal structure of the galaxy, likely stemming from a relatively recent gravitational interaction or gas accretion from a small companion that could have triggered the formation of the young star-forming regions observed along the entire galaxy.

\begin{figure}
    \includegraphics[width=\columnwidth]{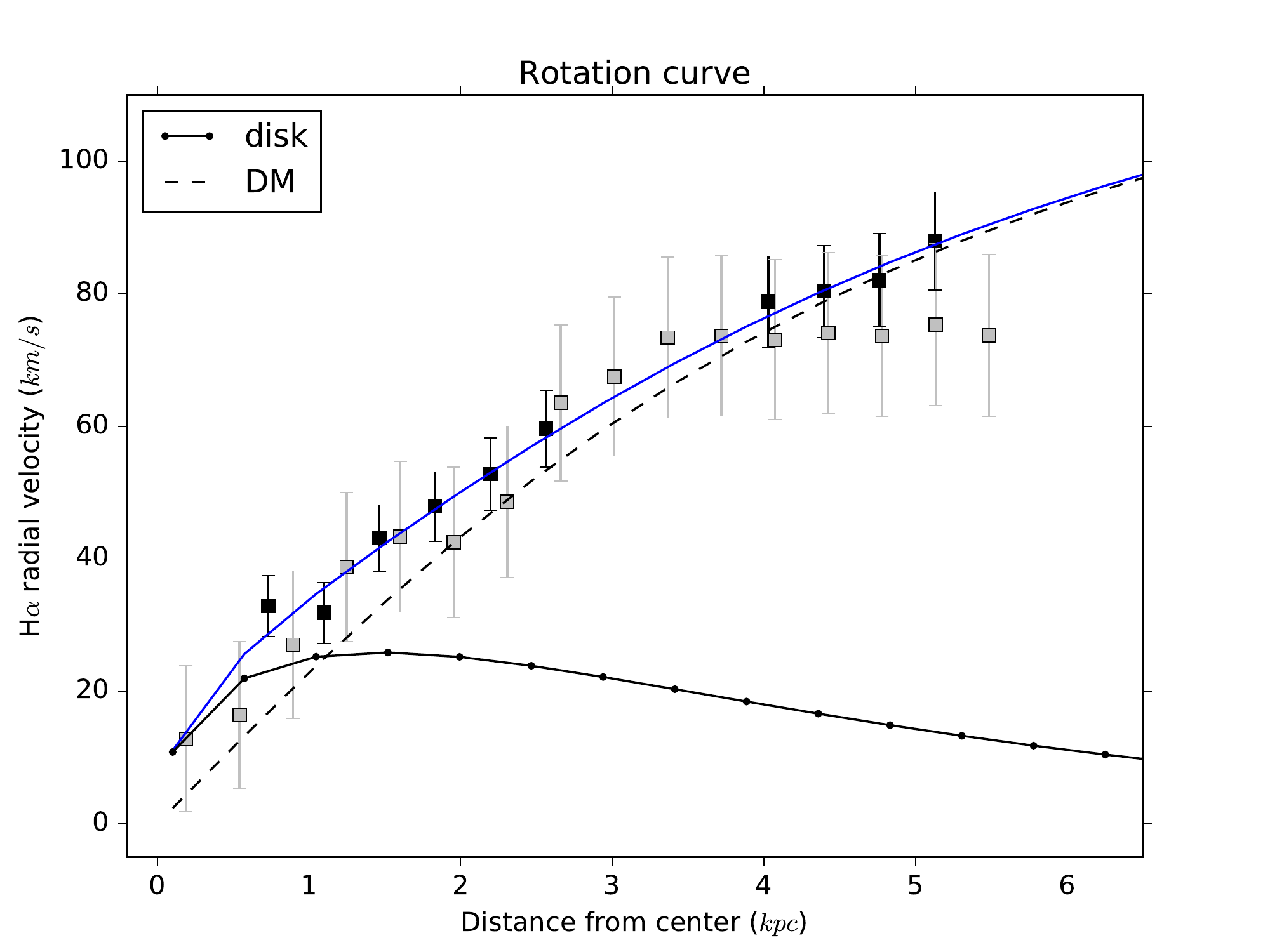}
    \caption{\emph{Left panel}: The best-fit (blue line) of the FLAMES/GIRAFFE (black squares) and MUSE (gray squares) velocity rotation curve  obtained using the model composed of 1) a pseudo-isothermal sphere due to DM (dashed line), and 2) a disk component (continuous black line). MUSE data are obtained from the average of the H$\alpha$ and the [\ion{O}{III}] 5007 \AA\, velocity rotation curve.}
   \label{fig:no10}
\end{figure}

Although the host galaxy of GRB 100316D is not a completely virialized system, we yet consider it to be of interest to fit the rotation curve of the GRB 100316D host galaxy, in order to compare our results with others published in the literature for similar galaxies. Typically, the rotation curves of dwarf irregular galaxies are fitted with a disk component, a gaseous component and a pseudo iso-thermal sphere due to dark matter (DM). The gas contribution is usually estimated from \ion{H}{I} 21 cm observations, but observations of this host galaxy led only to an upper limit on the \ion{H}{I} flux \citep{Michalowski2015}. Consequently, an additional input to rotation from the gas can not be excluded. We get a best fit with a disc mass-to-light ratio of $\Upsilon_{\odot} = 24.3 \pm 5.9$, a DM core radius of $r_{DM} = 4.0 \pm 0.9$ kpc and a central density of the DM halo, $\rho_{DM} = (29.9 \pm 5.6) \times 10^{-3}$~M$_{\odot}$/pc$^{-3}$, see Fig. \ref{fig:no10}. Early studies of late-type (Sd, Sm) spirals and dwarf irregular galaxies led to the evidence that these objects are entirely dominated by dark matter at all radii \citep{Broeils1992}, although more systematic analizes made on homogeneous samples show that these galaxies are not very different from brighter spirals \citep{Swaters2003}. The main differences compared to brighter spirals are the lower observed rotation velocities and the absence of a strong bulge component. Our analysis is in good agreement with this latter hypothesis: at smaller radii ($r \leq r_c$) the disk contribution dominates the rotation curve while at larger radii the DM dominates. Our results are also largely in agreement with other similar analyses of dwarf and low-surface-brightness galaxies \citep{Swaters2000,Swaters2003,Swaters2009}. In light of the results from the kinemetry analysis, which shows a system that is not completely virialized, we infer that caution is needed when conclusions based only on a single test are drawn. 

\section{Discussions and Conclusions}

The characteristics of MUSE are ideal to resolve and analize individual star-forming regions in low-redshift GRB host galaxies, such as the host of GRB~100316D. This galaxy is only the third GRB host that has been studied in high spatial resolution with IFU techniques (the other two being the host galaxy of GRB 980425 and GRB 060505, \citet{Christensen2008,Thoene2014,Kruhler2017}) and the second host of a GRB/SN. Studying individual star-forming regions in nearby host galaxies is of large importance when comparing them with the global properties of high-redshift host galaxies: as already reported in \citet{Christensen2008}, several uncertainties in the host properties would arise when the galaxy itself is unresolved. The host of GRB~100316D is a good example of this: the GRB location is exactly in the middle of two giant bright \ion{H}{II} regions whose integrated sSFRs and metallicities are at the extremes of their respective range of values (but within the calibration method's uncertainties, that are 0.16 and 0.18 dex, respectively). The metallicity (O3N2) at the GRB site is 12 + log(O/H) = $8.16 \pm 0.01$, in agreement with previous studies \citep{Starling2011}, which is based on the measured line fluxes, densities and temperatures. We also use two additional photo-ionisation-based metallicity indicators, the N2S2 index \citep{Dopita2016} and the semi-empirical estimator described in \citet{PerezMontero2014}, that provide quite different metallicity values for the GRB site: 12~+~log~(O/H)~=~7.99~$\pm$~0.01 and 12~+~log~(O/H)~=~8.42~$\pm$~0.01, respectively. Both these estimators show definitely different values with respect to the O3N2 and the N2 indices, but a similar offset between the N2S2 with respect the O3N2 indicator is observed for the GRB region in the host galaxy of GRB 980425  \citep{Kruhler2017}. In this case, the authors discuss that the origin of this observed discrepancy to the effects of the increasing ionization on the oxygen abundance parameter 12 + log(O/H). In the GRB 100316D host we equally observe a peak of the ionization parmeter $log U$ in correspondence of bright \ion{H}{II} regions.
Indeed, the measured metallicities with the O3N2 and N2 parameters in \ion{H}{II} regions, and in general in single spaxels, are slightly higher than the ones measured for the other two hosts analyzed with IFU data, see Fig. \ref{fig:no7}. Nevertheless, the values are still sub-solar (0.3 $Z_{\odot}$), as it is expected for GRB progenitor sites \citep{Fruchter2006}. 

For the SFR at the GRB site, we find 0.439 $\pm$ 0.001 M$_{\odot}$ yr$^{-1}$. \cite{Starling2011} report the value of 0.17 M$_{\odot}$ yr$^{-1}$ , which, however, refers to the \ion{H}{II}-3 region of the host galaxy for which we obtain 0.179 $\pm$ 0.001 M$_{\odot}$ yr$^{-1}$, in good agreement with their value. For the entire galaxy we find $SFR_{tot} = 1.20\pm0.08$ M$_{\odot}$ yr$^{-1}$ , not very different from the one provided by \citet{Michalowski2015}, $SFR = 1.73\pm0.08$ M$_{\odot}$ yr$^{-1}$, which however used radio observations for its estimate. The ionization parameter log\,U for the GRB region is $-2.74\pm0.01$, and is one of the lowest values found in GRB hosts: in the sample provided by \citet{Contini2017}, the median value for the ionization parameter is $\langle log U \rangle = -1.97$. We also find a density of the environment around the GRB location of $n_e \sim 100$ cm$^{-3}$ which is lower than typical values observed in other GRB hosts \citep{Savaglio2009}, but higher than the density values for the two bright nearby \ion{H}{II} regions. This result is in agreement with previous studies on the density surrounding GRB~100316D \citep{Starling2011}. 

We note that the measured extinction of the GRB location and the nearby star-forming regions is the lowest of the entire host galaxy, suggesting that the GRB region is not largely obscured by dust and consequently its location is prospectively favourable. The integrated host galaxy extinction we measure is $E(B-V) = 0.57 \pm 0.14$ mag, while for the GRB site we find a lower value of $E(B-V) = 0.42 \pm 0.01$ mag. In the literature slightly different values are reported for the host extinction: \citet{Starling2011} measure  $E(B-V) = 0.178$ mag from fitting of the spectral energy distribution of the \ion{H}{II}-3 region, \citet{Cano2011} find $E(B-V) = 0.18 \pm 0.08$ mag from the SN V-R colours at ten days from the SN peak and using the sample of \citet{Drout2011} as comparison, while \citet{Olivares2012} find $A_V = 1.2 \pm 0.1$ mag from X-ray to optical and NIR afterglow spectrum fitting, which translates to $E(B-V) = 0.27 \pm 0.02$ mag in the case of $R_V = 4.5$ \citep{Wiersema2011}. Finally, \citet{Michalowski2015} find $A_V = 0.86$ from an estimate of the UV attenuation and assuming a SMC extinction law, which implies $E(B-V) = 0.29$ mag. Comparing this result with a sample of dwarf galaxies with similar stellar mass from the SDSS DR7 \citep{Abazajian2009}, we find that for a given stellar mass, the maximum Balmer decrement would be $H\alpha/H\beta = 3.5$ \citep{Zahid2013}, whereas for the \ion{H}{II}-1 and the GRB region we have 3.8 and larger values for the other \ion{H}{II} regions. This implies that the extinction is actually higher than the typical values observed for galaxies in the SDSS DR7 sample. 

The host galaxy of GRB~100316D is a perturbed late-type dwarf galaxy at moderate inclination. Although from the rotation curve the galaxy does not show strong signs of  interactions/merging, the kinemetry analysis suggests a system out of equilibrium: from the $k5/k1$ ratio we infer a highly unstable disk and the presence of possible multiple kinematic components. A peak in the distribution of the $k5/k1$ ratio corresponds to the bright star-forming system around the \ion{H}{II}-1-3 regions, where the GRB was located. We also find high $k5/k1$ values at a position near the center of the galaxy, i.e. where we observe an enhancement in the H$\alpha$-FWHM map and where we find shock signatures. The N2 map also shows a higher metallicity corresponding to this central structure, see Fig. \ref{fig:no4}. The possibility of a relatively recent gravitational encounter finds support in other observational evidences: the LINER-like region located in the region of the host with large H$\alpha$ dispersion velocities can also be attributed to an on-going cloud collision  powered by a recent interaction with some small companion giving rise to thermal shocks. We further note that \citet{Michalowski2015} reports the detection of a faint \ion{H}{I} structure which intersects the galaxy disk at exactly the location of this central region, while for the rest of the galaxy no signal in radio has been reported. This possibility would also explain the perturbed morphology of the galaxy and the presence of peripheral bright star-forming regions : the brightest star-forming regions, and the GRB position itself, are located at the edge of this large structure. This gravitational encounter could have compressed the gas a few million of years ago and then triggered an intense star-formation period in regions with low-metallicity environments. 

Whatever is the cause of the intense star formation, it could have permitted the formation of massive runaway stars also in local clusters \citep{Billett2002}, and possibly to the presence of high-mass runaway stars \citep{Hammer2006}, one if which could be the progenitor of GRB 100316D that would anyway remain quite close to their birthplace given their limited lifetime \citep{Crowther2013}, explaining the offset between the explosion site and the peak of HII region. From the estimated age (20--30 Myr) of the GRB progenitor star and from the projected distance, $d = 660$ pc, between the GRB location and the peak of the \ion{H}{II}-1 region, we estimate a possible kick velocity for the GRB progenitor of 30 km/s, which is actually a lower limit, namely the distance $d$ projected onto the perpendicular plane. This value is in line with maximum kick velocities expected for runaway stars, ejected during the process of star-formation or in the case of a binary-system SN explosion \citep{Fujii2011,Crowther2013}. Another possibility could be that the \ion{H}{II} region that formed the GRB progenitor has been diluted, as the typical age of large star-forming regions is similar to the age estimated for the GRB progenitor. This scenario finds support also in the observed metallicity values at the GRB site and \ion{H}{II}-1 region: GRB progenitors mainly form in low-metallicity environments \citep[][ and references therein]{Graham2017}, and although the GRB site is characterized by a low metallicity  ($Z_{GRB} = 0.3 Z_{\odot}$), the nearby \ion{H}{II}-1 region shows even lower values. Consequently, we can not rule out the possibility that the GRB progenitor has been ejected from its original formation site by an energetic physical mechanism, as proposed above.

The presence of the [\ion{N}{I}] 5200 \AA\, line in the spectra of the \ion{H}{II}-1,2 regions and  the GRB site provides important clues on the stellar population and GRB progenitor properties: its origin is the result of photon pumping by ultraviolet radiation coming from the massive stellar population embedded in the \ion{H}{II} region. The ratio of the observed fluxes $I([\ion{N}{I}])/I(H\beta)$ suggests late O--early B spectral type for this stellar population, in analogy with results obtained from a detailed analysis of the Orion nebula system \citep{Baldwin2000,Ferland2012}, characterized by masses ranging from 20 and 40 $M_{\odot}$ \citep[][ see also \citealp{Bastian2010} for a general review]{Hohle2010}. A similar GRB progenitor mass can been inferred from the estimate of the stellar age, ranging from 20 to 30 Myr, using different line indicators for the GRB site spectrum. This value is also in line with the value proposed by \citet{Starling2011} where the authors used the observed $H\delta$ absorption feature due to the underlying stellar component. 

With the increasing number and capabilities of IFU instruments, both from ground-based telescopes and in space with the upcoming NIRSpec on-board the James Webb Space Telescope, resolved studies of higher-redshift GRB host galaxies will soon be possible. The uncertainties in the evolution with time of the gas properties and the star-formation history in different kinds of GRB hosts will be better quantified by studying these galaxies with IFU techniques. In the host of GRB~100316D, we clearly see that the explosion did not happen in an average region of the galaxy. Integrated measurements obtained from long-slit spectra usually underestimate the extreme nature of GRB explosion sites \citep{Christensen2008,Thoene2014,Graham2017}, as we have noted for the extinction, the EW values for Balmer lines and the ionization parameter, whose exact knowledge is critical for understanding GRB progenitor models. At the same time, broad-band imaging has revealed a metallicity bias that is a bit below solar metallicity \citep{Fruchter2006,Larsson2007,Raskin2008,Perley2015,Schulze2015,Vergani2017}. However, these latter measurements are the result of a spatial integration on the entire host-galaxy emission, while IFU observations in a single shot could provide critical constraints on the true metallicity bias and identify other effects, like internal perturbations (giving rise to bursts of star-formation), and then providing additional information on the initial mass function and its variation with redshift.

\section*{Acknowledgements}

We thank the referee for her/his scientific comments that allowed to improve the analysis as well as the current manuscript.
LI, CT, ZC, AdUP and DAK acknowledge support from the Spanish research project AYA 2014-58381-P. CT and AdUP furthermore acknowledge support from Ram\'on y Cajal fellowships RyC-2012-09984 and RyC-2012-09975. DAK and ZC acknowledge support from Juan de la Cierva Incorporaci\'on fellowships IJCI-2015-26153 and IJCI-2014-21669. RSR is supported by a 2016 BBVA Foundation Grant for Researchers and Cultural Creators. SS acknowledges support from the CONICYT Chile FONDECYT Postdoctorado fellowship 3140534 and the Feinberg Graduate School. SS and FEB acknowledges support from CONICYT-Chile  (Basal-CATA PFB-06/2007, FONDECYT Regular 1141218), the Ministry of Economy, Development, and Tourism's Millennium Science Initiative through grant IC120009, awarded to The Millennium Institute of Astrophysics, MAS. RA acknowledges the support from the ERC Advanced Grant 695671 'QUENCH'. PJ acknowledges support by a Project Grant (162948 - 051) from The Icelandic Research Fund. SDV acknowledges the support of the French National Research Agency (grant ANR-16-CE31-0003). SDV acknowledges the support of the French National Research Agency (grant ANR-16-CE31-0003). AR acknowledge support from Premiale LBT 2013.




\bibliographystyle{mnras}





\begin{figure*}
	\includegraphics[width=2\columnwidth]{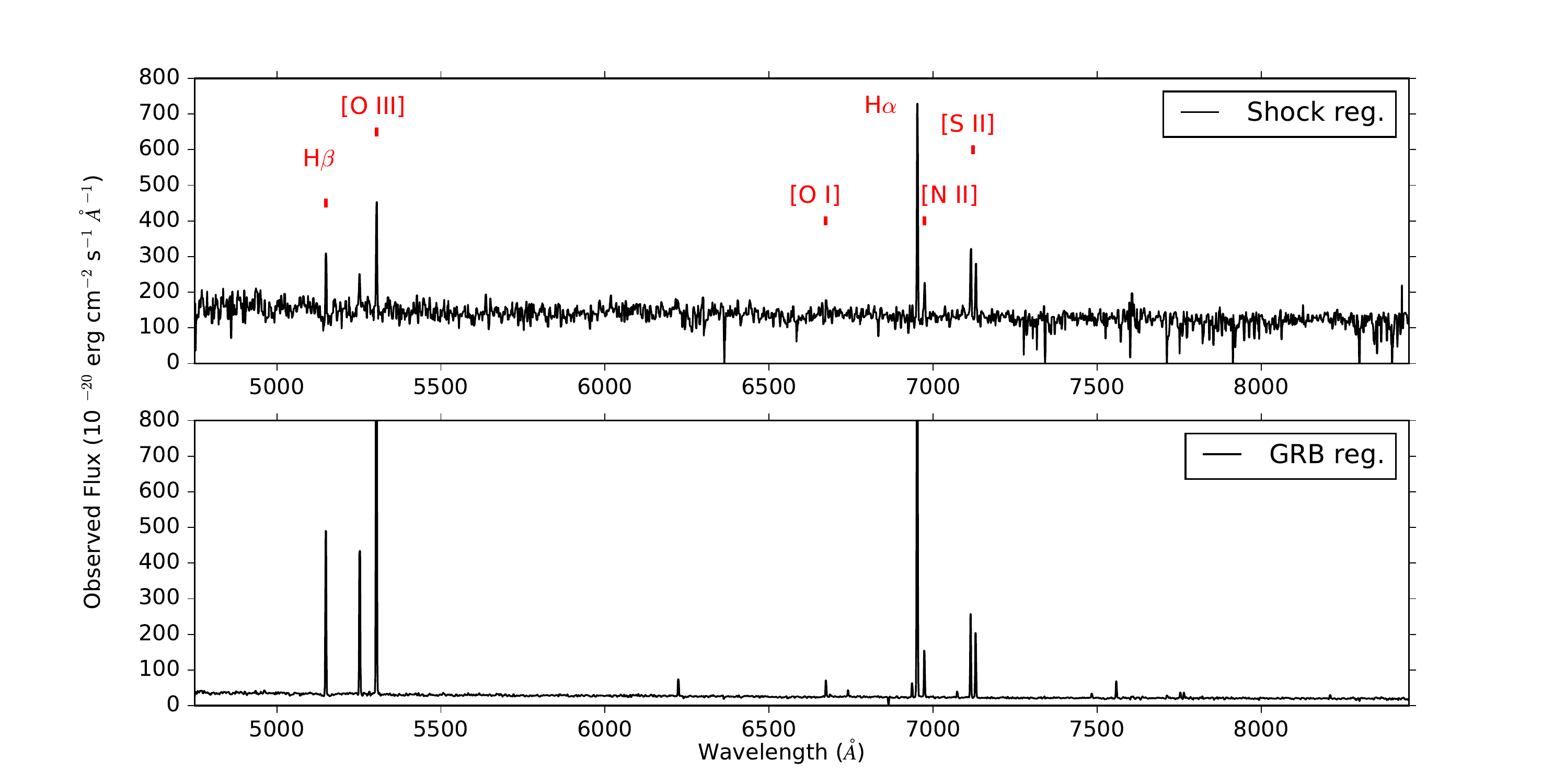}
    \caption{\textit{Upper panel}: Spectrum of the possibly shocked region showing a high N2 metallicity and a larger FWHM of the emission lines. This region is referred to in the BPT diagram as black triangle and the label ''shocked region''. Note the presence of 
bright [$\ion{N}{II}]\lambda6584$ and [$\ion{S}{II}]\lambda\lambda6717/32$ emission lines, compared to the 
strength of H$\beta$. \textit{Lower panel}: The spectrum of the GRB 
region as a comparison.}
    \label{fig:app2}
\end{figure*}

\begin{landscape}
\begin{table}
	\centering
	\caption{Extinction-corrected fluxes for the corresponding emission lines in units of 10$^{-18}$ erg/cm$^2$/s. We do not observe any of the lines reported in the table in the  \ion{H}{II}-7 and \ion{H}{II}-8 regions.}
	\label{tab:app1}
	\begin{tabular}{lccccccccccc}
\hline
       & \ion{He}{I}            & \ion{He}{I}             & \ion{He}{I}           & \ion{He}{I}             & [\ion{N}{I}]              & [\ion{Ar}{III}]          & [\ion{Ar}{III}]         & [\ion{S}{III}]            & [\ion{Fe}{III}]          & [\ion{Fe}{III}]         \\
              & $\lambda$ 4912            & $\lambda$ 5016             & $\lambda$ 6678           & $\lambda$ 7066             & $\lambda$ 5200              & $\lambda$ 7136          & $\lambda$ 7751         & $\lambda$ 6312            & $\lambda$ 4659          & $\lambda$ 4987         \\
\hline
 HII-1 & 12.53$\pm$5.58  & 34.26$\pm$5.74 & 72.72$\pm$5.18 & 51.481$\pm$4.89  & 20.25$\pm$5.47 & 169.85$\pm$5.77 & 84.26$\pm$4.66 & 56.18$\pm$5.04 & 26.39$\pm$5.89 & 19.26$\pm$5.58 \\
 HII-2 & 2.31$\pm$7.09   & - & 42.35$\pm$6.54 & 24.782$\pm$6.238 & 12.87$\pm$6.98 & 125.28$\pm$6.79 & 51.44$\pm$6.04 & 36.71$\pm$6.58 & 20.13$\pm$7.48 & 18.78$\pm$7.16 \\
 HII-3 & 1.96$\pm$3.9    & 13.28$\pm$4.04 & 26.95$\pm$3.58 & 21.394$\pm$3.377 & 8.41$\pm$3.85  & 70.57$\pm$3.86  & 32.17$\pm$3.21 & 22.9$\pm$3.51  & 4.37$\pm$4.16  & 6.15$\pm$3.93  \\
 HII-4 & - & 10.55$\pm$5.89 & 12.92$\pm$6.41 & 3.507$\pm$6.448  & - & 26.23$\pm$6.6   & 12.29$\pm$6.38 & 9.69$\pm$6.44  & - & - \\
 HII-5 & -  & 23.56$\pm$6.1  & 9.59$\pm$2.79  & 13.869$\pm$5.645 & 3.98$\pm$5.82  & 44.17$\pm$5.87  & 36.75$\pm$5.61 & 5.34$\pm$3.02 & - & - \\
 HII-6 & 4.26$\pm$1.92   & 3.43$\pm$1.86  & 4.39$\pm$1.72  & 3.288$\pm$1.701  & -  & -  & - & -  & - & -  \\
 HII-7 & - & -  & -  & -  & -  & -    & -  & -  & - & -  \\
 HII-8 & - & -  & -  & -  & -  & -    & -  & -  & - & -  \\
 GRB   & -    & 1.93$\pm$1.28  & 2.63$\pm$1.16  & 1.09$\pm$0.77  & -  & 5.29$\pm$1.25   & -  & 1.07$\pm$0.96  & - & 1.84$\pm$0.96 \\
\hline
\end{tabular}
\end{table} 
\end{landscape}




\bsp	
\label{lastpage}
\end{document}